\documentclass[vecphys]{svmult}		% vectors in bold face
\usepackage{makeidx}         % allows index generation
\usepackage{graphicx}        % standard LaTeX graphics tool
\usepackage{epsfig}          % another way to include figures
\usepackage{multicol}        % used for the two-column index
\usepackage[bottom]{footmisc}% places footnotes at page bottom

\usepackage{times}
\usepackage{amsmath}
\usepackage{amssymb}

\usepackage{color}

\makeindex             % used for the subject index
\begin{document}
%%%%%%%%%%%%%%%%%%%%%%%%%%%%%%%%%%%%%%%%%%%%%%%%%%%%%%%%%%%%%%%%%%%%%

\def\jcmindex#1{\index{#1}}
\def\myidxeffect#1{{\bf\large #1}}
\def\jcm#1{{\color{blue} #1}}

\newcommand{\wb}{\omega_\mathrm{b}}
\newcommand{\wo}{\omega_\mathrm{o}}
\renewcommand{\wp}{\omega_\mathrm{ph}}
\newcommand{\e}{\mathrm{e}}

% Title
\title*{Discrete breathers in $\phi^4$ and related models}
\titlerunning{Discrete breathers in $\phi^4$  and related models}
\author{
Jes\'us Cuevas--Maraver%\inst{1,2}
\and
Panayotis G. Kevrekidis%\inst{3}
}

\institute{
Jes\'us Cuevas--Maraver
\at
Grupo de F\'{\i}sica No Lineal, Universidad de Sevilla, Departamento de F\'{i}sica Aplicada I, Escuela Polit\'ecnica Superior.
C/ Virgen de \'{A}frica, 7, 41011-Sevilla, Spain,
\and
Jes\'us Cuevas--Maraver
\at
Instituto de Matem\'aticas de la Universidad de Sevilla (IMUS). Edificio Celestino Mutis. Avda. Reina Mercedes s/n, 41012-Sevilla, Spain
\email{jcuevas@us.es}
\and
Panayotis G. Kevrekidis
\at
Department of Mathematics and Statistics, University of Massachusetts, Amherst, MA 01003-4515, USA
\email{kevrekid@math.umass.edu}
}
\maketitle
\abstract
    {In this Chapter, we touch upon the wide topic of discrete breather (DB)
      formation with a special emphasis on the prototypical system
      of interest, namely the $\phi^4$ model. We start by introducing the
      model and discussing some of the application areas/motivational
      aspects of exploring time periodic, spatially localized
      structures, such as the DBs. Our main emphasis is on the existence,
      and especially on the stability features of such solutions.
      We explore their spectral stability numerically, as well
      as in special limits (such as the vicinity of the so-called
      anti-continuum limit of vanishing coupling) analytically.
      We also provide and explore a simple, yet powerful stability
      criterion involving the sign of the derivative of the
      energy vs. frequency dependence of such solutions. We then
      turn our attention to nonlinear stability, bringing forth
      the importance of a topological notion, namely the Krein signature.
      Furthermore, we briefly touch upon linearly and nonlinearly unstable
      dynamics of such states. Some special aspects/extensions of such
      structures are only touched upon, including moving breathers and
      dissipative variations of the model and some possibilities for
      future work are highlighted. While this Chapter by no means
      aspires to be comprehensive, we hope that it provides some
      recent developments (a large fraction of which is not
      included in time-honored DB reviews) and associated future
      possibilities.
}

\section{A brief description of discrete breathers: definition, historical perspective and applications}

Dynamics of localized excitations is, undoubtedly, one of the most important topics within the realm of nonlinear science. In discrete systems, the localized excitations that can be argued to be most generic~\cite{Flach1,Flach2,Aubry2}
are the so-called discrete breathers (DBs). These can be defined as time-periodic spatially-localized coherent structures emerging at coupled nonlinear
oscillator lattices. This term was coined by Campbell and Peyrard \cite{CP} in order to distinguish them from the ``continuous'' breathers found
as exact solutions through the inverse scattering
machinery in the sine-Gordon PDE \cite{AKNS}.
[It is worthwhile to mention in passing that continuous breathers have
  a particularly interesting history associated with them in
  $\phi^4$ models~\cite{Kruskal,Boyd},
  which is described in detail in a different Chapter
  of this special volume.]
Such DBs are also known as intrinsic localized modes; this name was introduced
in order to emphasize that their origin was in the intrinsic nonlinearity of the system, contrary to the case of Anderson modes, where localization stems from lattice disorder~\cite{Ander} and can exist even in the linear limit.

A major development that springboarded the study of
DBs took place in 1988, through the numerical study of
DBs in some prototypical lattice models as
reported in
two pioneering works \cite{ST,TKS} by Takeno, Sievers and Kisoda. In these papers DBs are calculated for the first time in the two basic kinds of lattices where they can emerge, namely Fermi--Pasta--Ulam--Tsingou (FPUT)\footnote{These lattices have been traditionally denoted as Fermi--Pasta--Ulam, forgetting the outstanding role of Mary Tsingou who was the person responsible for
  all the numerical simulations in these first computations of nonlinear lattice dynamics \cite{Tsingou}} \cite{ST} and Klein--Gordon (KG) \cite{TKS}.
Additionally, some of their stability properties were also
determined. FPUT lattices are characterized by the absence of substrate (on-site) potential and the nonlinearity of the inter-site forces, whereas KG lattices possess nonlinear on-site potential and (typically) linear inter-site forces.
Interestingly, in those references, potentials of the $\phi^4$ form and related models were considered. There are also mixed cases of lattices where nonlinearities are present in both the substrate and intersite potentials (see e.g. \cite{kimura});
these lattices are sometimes denoted as KG/FPUT lattices.

While these early works provided credible numerical evidence and
planted the seed for studying DBs, they
did not rigorously prove their existence. The latter
came in a celebrated 1994 paper by MacKay and Aubry \cite{MA} and was based on the concept of the so-called anti-continuous (AC) limit. They basically demonstrate in a rigorous fashion, by making use of the implicit function theorem, that DBs can generically exist in nonlinear KG lattices as a result of
unobstructed (when some suitable resonance restrictions
are avoided) continuation of periodic orbits of individual (uncoupled)
oscillators. Mackay--Aubry's theorem, together with the rigorous proof of stability by Aubry in 1997 \cite{Aubry} and the numerical methods developed by Mar\'{\i}n among others \cite{Marin} led to an intense interest on DBs in the late 1990's and the beginning of the 21st Century, not only from the theoretical but also from the experimental point of view. The
relevant activity has been very well summarized
by now in a series of reviews; see,
e.g.,~\cite{Flach1,Flach2,Aubry2,QiM,Velarde}.

DBs have been sought for in many fields of Physics. They have been experimentally generated in arrays of Josephson junctions \cite{Trias,Ustinov}, mechanical \cite{Lars} and magnetic pendula \cite{Russell2} and microcantilevers \cite{Sato}, nonlinear electrical lattices \cite{Faustino1,Faustino2} and granular media \cite{granular2,granular}. They have been observed in molecular and ionic crystals like the so-called PtCl \cite{Swanson} and antiferromagnets \cite{Sato2}, and have been
argued as plausible explanations of observational
findings in systems such as $\alpha$-Uranium \cite{Uranium} or NaI \cite{NaI}.
They have also been speculated to play an important role in DNA denaturation
in an extensive literature reviewed, e.g., in~\cite{Peyrard},
as well as to arise in reconstructive chemical reactions~\cite{Mica} and in the slow decay of luminiscence in Pb-doped alkali-halide crystals as KBr \cite{Mihokova}, and predicted to exist in crystals of Niobium and Nickel \cite{Niobium} or carbon materials like graphene, carbon nanotubes, fullerenes or hydrocarbons (see \cite{graphene} for a review). Under suitable conditions, DBs can move along the lattice and are known in that setting as moving breathers.
As such, they have been argued to be responsible  for the
observed tracks in muscovite mica sheets \cite{QiM,Juan1} and of the infinite charge mobility (also dubbed as \emph{hyperconductivity}) experimentally evidenced in such crystals \cite{Russell1,Juan2}. Admittedly, these are only
some examples of an ever increasing list of applications which is
by necessity, due to the limited scope of this Chapter, rather incomplete.
Nevertheless, it serves to illustrate the generic nature and wide
impact of such structures and their broad relevance of study.

The present Chapter is devoted to reviewing more concretely
some results on the existence, stability and dynamics
of DBs in KG lattices with $\phi^4$ on-site potential, and some other miscellaneous topics related to DBs in such lattices; many of these results have been found for generic KG lattices, but we will focus here on the principal
theme of this special volume, namely the $\phi^4$ potential.
Moreover, by choice, the emphasis will be on some recent results, not
only due to their connections to the research interests and recent
work of the authors, but also because, to the best of our knowledge,
the associated findings and the resulting over-arching
stability perspective have not been collected in such a summarizing
body of work to date elsewhere.

\section{The Klein-Gordon lattice and the anti-continuous limit}
\label{sec:AClimit}

After the brief presentation of the DB concept, in this Section we will introduce some definitions. First of all, we need to address the concept of
Klein-Gordon nonlinear dynamical lattice; from a mathematical point of view, it can be defined as a system of coupled second-order ordinary differential equations of the form:

\begin{equation}\label{eq:dyn0}
    F_n(u)\equiv\ddot{u}_n+V'(u_n)+\sum_m C_m (u_n-u_{n+m}-u_{n-m})=0
\end{equation}
where $n$ and $m$ are $D$-dimensional indices, $V(u_n)$ is the on-site potential, which is not necessarily homogeneous, and $C_m$ is the coupling constant, depending on the distance from the neighbors. In most case examples, the inter-site force is nearest-neighbour with $C_m=C\delta_{m,1}$, in which case the
previous equation can be written e.g. for the one-dimensional lattice as
\begin{equation}\label{eq:dyn1d}
    F_n(u)\equiv\ddot{u}_n+V'(u_n)+C(2 u_n-u_{n+1}-u_{n-1})=0.
\end{equation}
This dynamical equation derives from the following Hamiltonian:

\begin{equation}\label{eq:ham1d}
    H=\sum_n h_n=\sum_n \frac{1}{2}\dot{u}_n^2+V(u_n)+\frac{C}{4}\left[(u_n-u_{n-1})^2+(u_n-u_{n+1})^2\right]
\end{equation}
with $h_n$ being the energy density. As mentioned above, the anti-continuous (AC) limit introduced by MacKay--Aubry's theorem \cite{MA} is of paramount importance in order not only to prove the existence of discrete breathers in nonlinear KG lattices, but also to give a hint on how to obtain them numerically.
The AC limit is that of all the lattice oscillators being
uncoupled (i.e. $C=0$) and either oscillating with the same frequency $\wb$ or remaining at rest. By virtue of the implicit function theorem, MacKay and Aubry established that the solution can be continued from the AC limit to a finite value of $C\rightarrow0$ whenever two conditions are fulfilled: (1) the potential of the isolated oscillators is anharmonic and (2) no integer multiples of
the DB frequency $\wb$ resonate with the linear modes frequency (i.e. the
so-called phonon band which we will quantify further below).
If these conditions are fulfilled, a coherent (in the sense that all the sites oscillates with the same frequency) and exponentially localized structure
of a DB form will exist. These discrete breathers are exact periodic
orbit solutions (up to machine precision) of the nonlinear KG equation.

This theorem is not only of theoretical value, but also of practical
usefulness as it provides a strategy on how
to numerically calculate DBs that was exploited by Mar\'{\i}n and collaborators \cite{Marin}. Methods based on the AC limit are quite simple: as at the AC limit the oscillators are uncoupled, it suffices to get a periodic orbit of single oscillators subjected to potential $V(u)$ and continue this solution by means of fixed-point methods (like Newton-Raphson) up to the desired coupling. Among the numerical methods used for attaining DBs, we can highlight two: (1) Fourier space methods in which DBs are represented by a Galerkin truncation up
to index $k_m$ in a Fourier series expansion of the form:
\begin{equation}\label{eq:Fourier}
    u_n=\sum_{k=-k_m}^{k_m} z_k \e^{\mathrm{i}k\wb t}\ ,
\end{equation}
transforming the coupled ODE system (\ref{eq:dyn0}) into a set of nonlinear algebraic equations; and (2) shooting methods, where DBs are fixed points of the map:
\begin{equation}\label{eq:shooting}
    \left(\{u_n(0)\},\{\dot{u}_n(0)\}\right)\rightarrow\left(\{u_n(T)\},\{\dot{u}_n(T)\}\right)\ ,
\end{equation}
with $T=2\pi/\wb$ being the breather period. Fourier methods have the advantage of using an analytical Jacobian but one has to pay the price of handling with a larger number of equations. In addition, there are pathological potentials \cite{Faustino1,Hertzian} for which the convergence of Fourier series is very slow and this method cannot be used.

There is a great number of potentials that can be found in the DB literature.
They  can be classified as soft or hard, if the energy decreases or increases, respectively, with the frequency; a simple way to discern if a potential is soft or hard is by making use of the hardening coefficient $h=3V''''(0)-5V'''(0)$ \cite{Guillaume} so that the potential is soft (hard) when $h<0$ ($h>0$). Moreover, a soft (hard) oscillator vibrates with a frequency $\wb$ which is smaller (greater) than its natural frequency $\wo=\sqrt{V''(0)}$ (notice that in most cases, $\wo=1$). Typical examples
of soft potentials include the Morse, Lennard-Jones, cubic ($\phi^3$), sine-Gordon and double-well potentials, many of which arise in applications~\cite{Flach1,Flach2}. Hard potentials are usually particular cases of polynomials potential which, depending of parameters, can be either soft or hard, namely cubic-quartic or purely quartic ($\phi^4$) anharmonicities, where the
quartic term arises with a positive sign; see below.
This latter potential, on which we will focus in the present Chapter,
is given by
\begin{equation}\label{eq:phi4potential}
  V(u)=\frac{1}{2}u^2+\frac{1}{4}su^4
\end{equation}

When $s=1$ ($s=-1$), the potential is hard (soft). The orbits $u(t)$ of an isolated oscillator can be expressed in terms of Jacobi elliptic functions and the modulus $m$ is related to the oscillation frequency $\wb$ through a transcendental equation. Thus, if $s=1$,
\begin{equation}\label{eq:1oschard}
    u(t)=\sqrt{\frac{2m}{1-2m}}\mathrm{cn}\left(\frac{2K(m)\wb t}{\pi},m\right), \qquad \wb=\frac{\pi}{2\sqrt{1-2m}K(m)}
\end{equation}
and, if $s=-1$
\begin{equation}\label{eq:1oscsoft}
    u(t)=\sqrt{\frac{2m}{1+m^2}}\mathrm{cd}\left(\frac{2K(m)\wb t}{\pi},m\right), \qquad \wb=\frac{\pi}{2\sqrt{1+m^2}K(m)}
\end{equation}

\begin{figure}[tbp]
\begin{tabular}{cc}
\includegraphics[height=4.5cm]{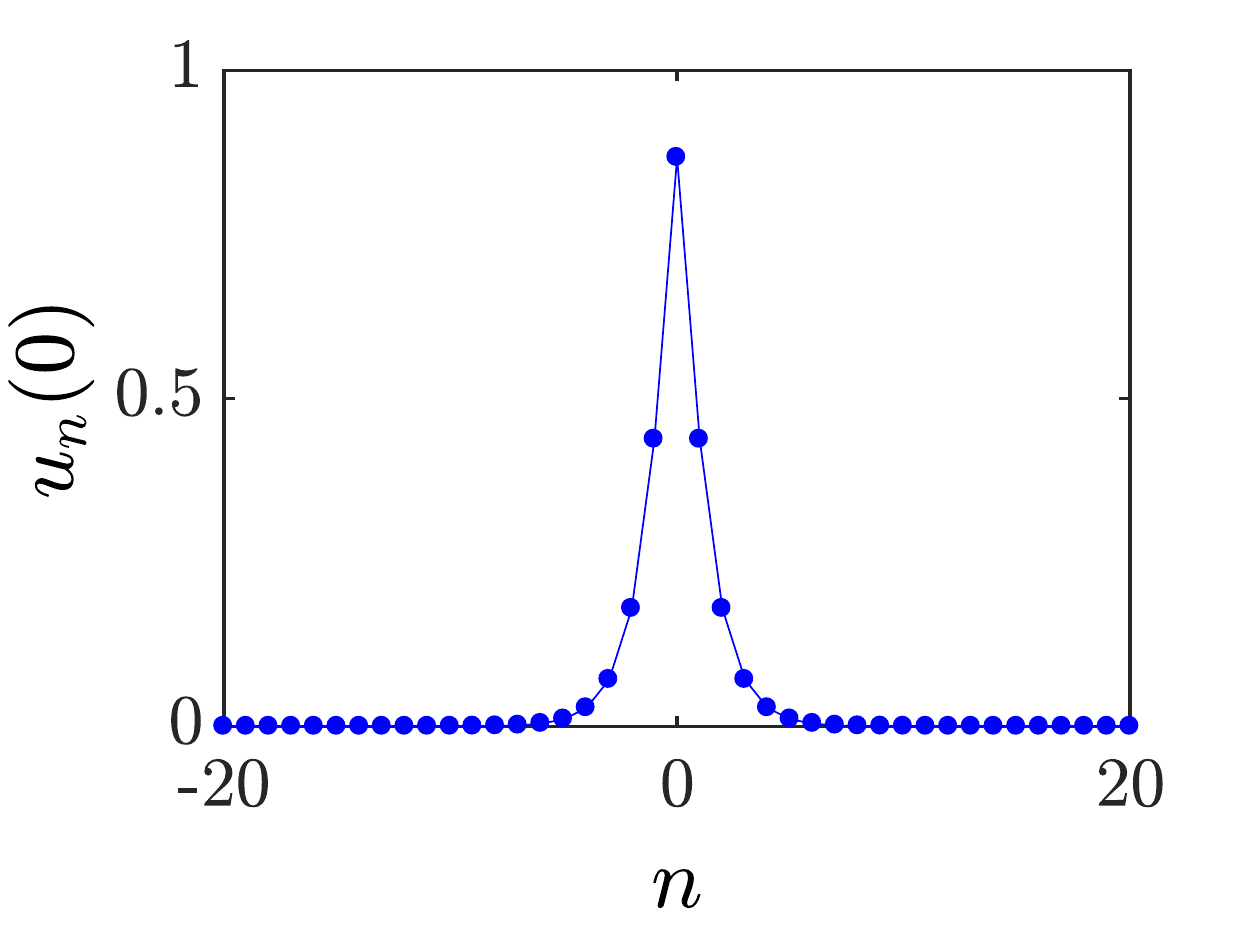} &
\includegraphics[height=4.5cm]{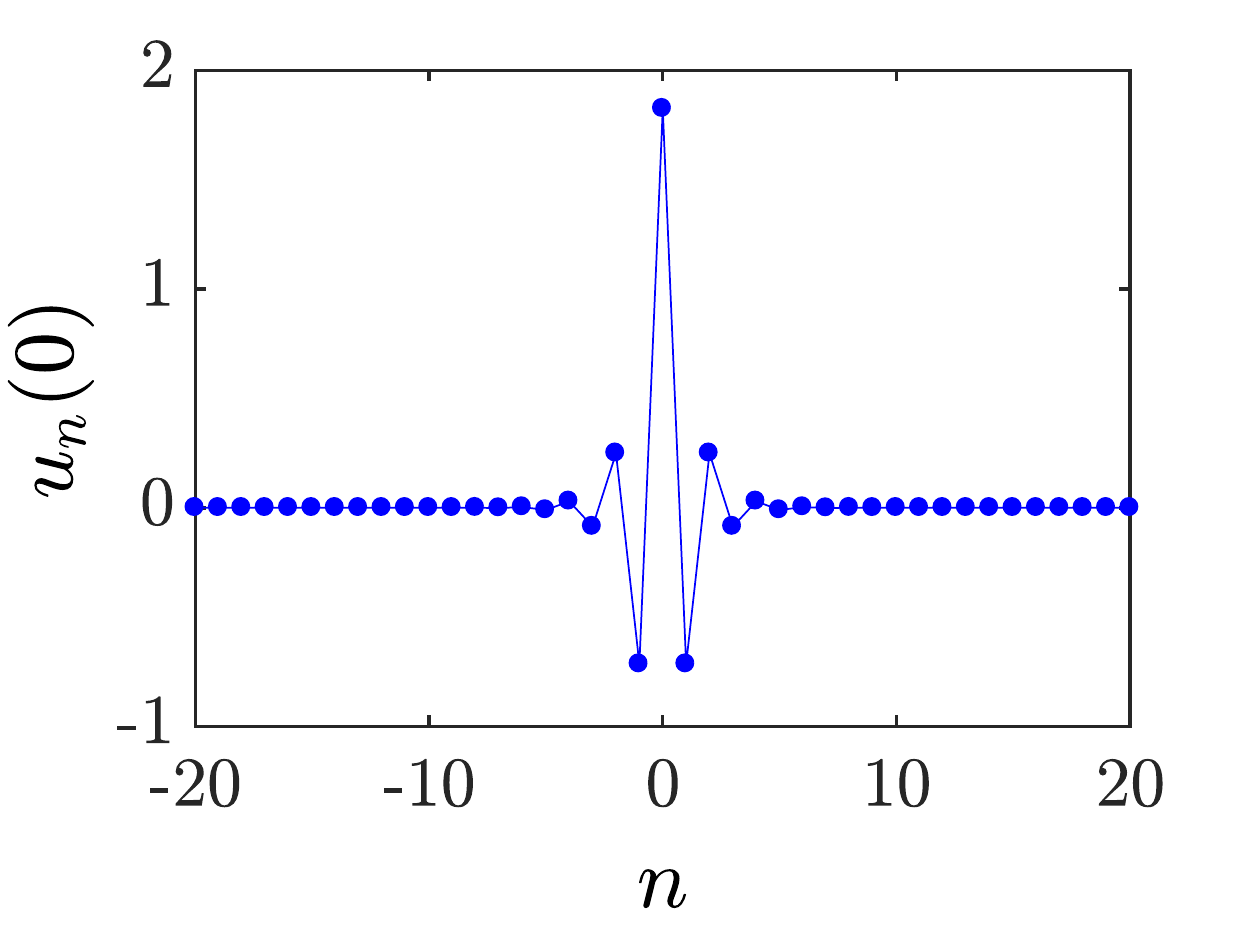} \\[3mm]
\includegraphics[height=4.5cm]{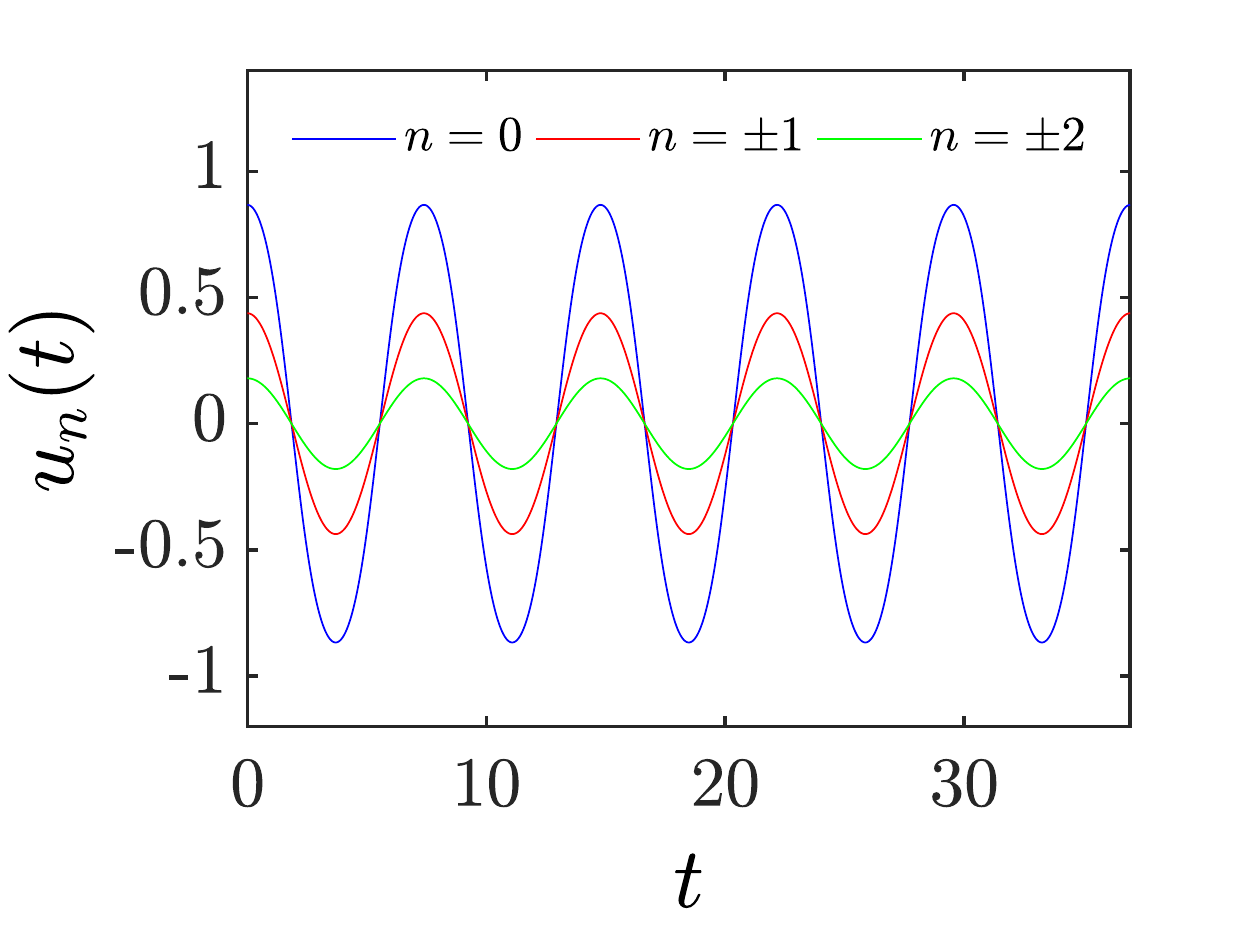} &
\includegraphics[height=4.5cm]{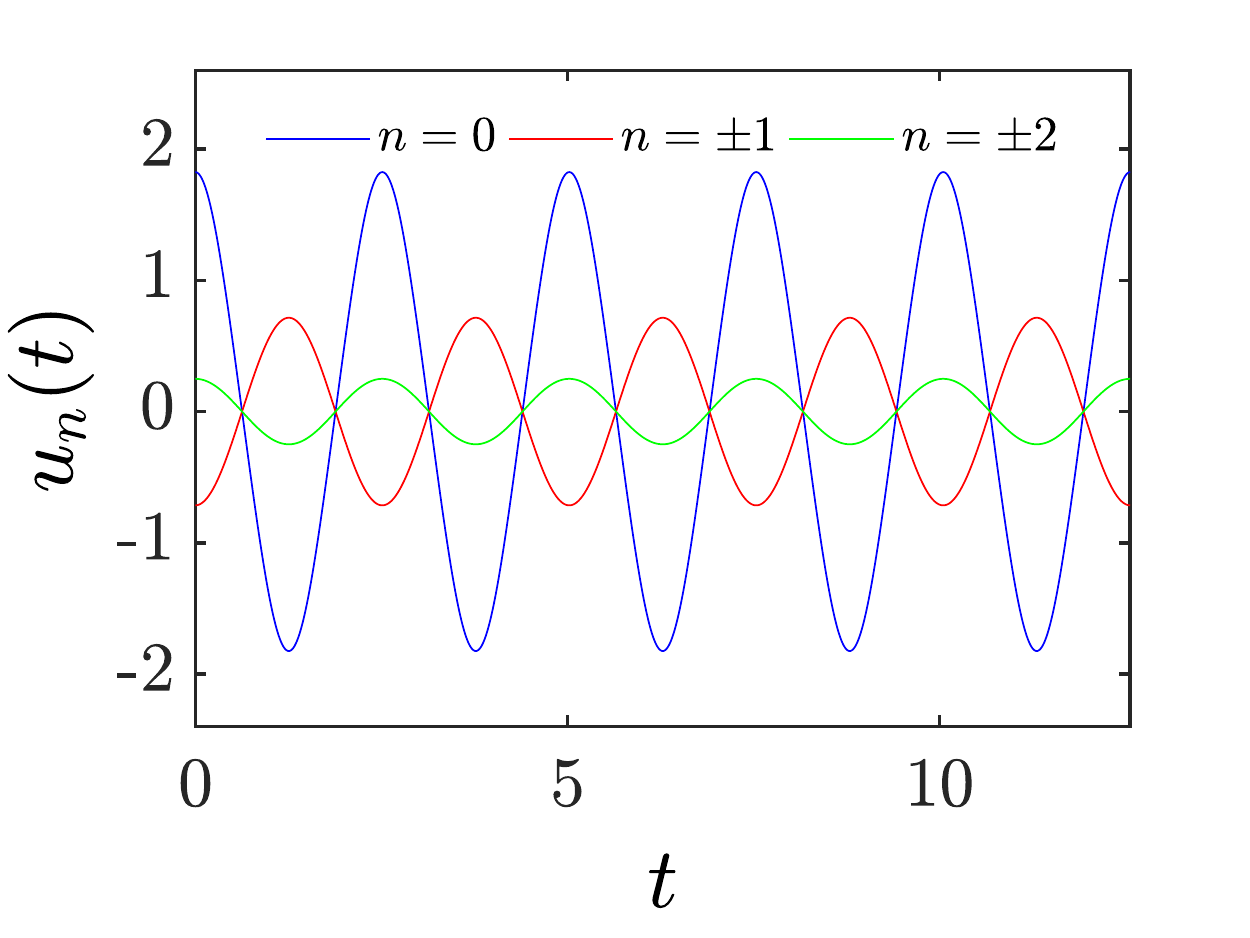} \\[3mm]
\includegraphics[height=4.5cm]{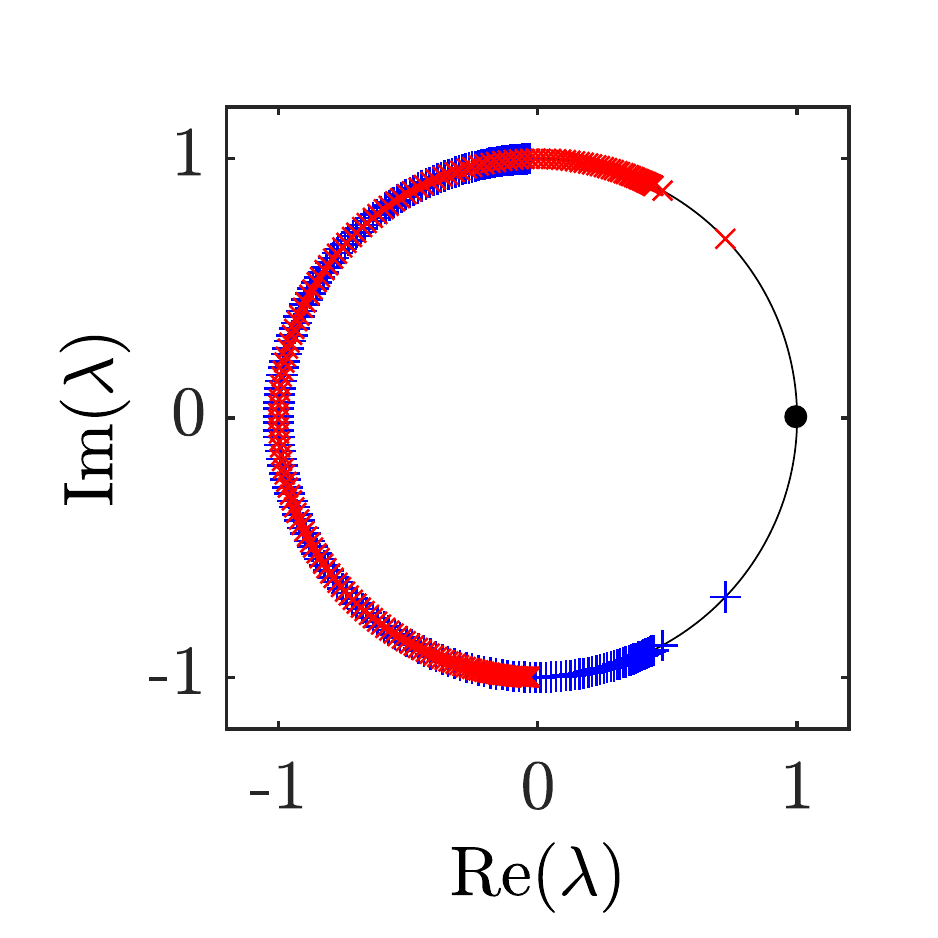} &
\includegraphics[height=4.5cm]{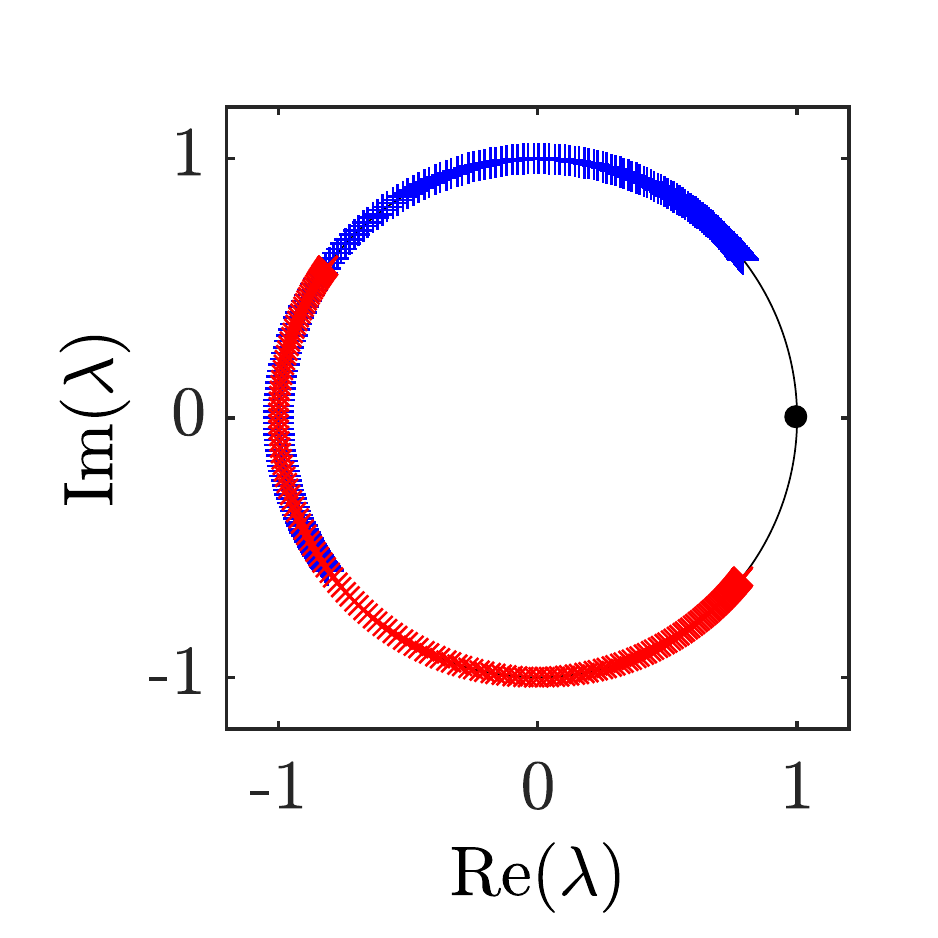} \\
\end{tabular}
\caption{Profile (top panels), time-dependence of the central sites (middle panels) and Floquet multiplier spectrum (bottom panels) of a 1-site breather in a soft (left panels) and in a hard (right panels) potential. In the former case, parameters are $\omega_\mathrm{b}=0.85$ and $C=0.3$, whereas in the latter case, $\omega_\mathrm{b}=2.5$ and $C=1$. In the right panels, multipliers with positive (negative) Krein signatures are depicted with red crosses $\times$ (blue pluses $+$); see the associated discussion around Eq.~(\ref{eq:Krein}) below.}
\label{fig:example}
\end{figure}

Notice that in the soft case, there are heteroclinic orbits separating oscillating states from unbounded ones representing escape from the potential \cite{Vassos}. Another interesting feature that distinguishes soft and hard potentials is related to the oscillation pattern of the tails. When the potential is hard, the tails are staggered and they are unstaggered in the soft case~\cite{Chen}.
An example of discrete breathers in soft and hard potentials is shown in Fig.~\ref{fig:example} where the profile and the time-evolution is displayed.

At the AC limit, it is possible to construct DB-like solutions with more than one excited sites, dubbed as {\em multibreathers}. When suitably constructed
(typically with the ``excited'' oscillators in- or out-of-phase
as summarized, e.g., in~\cite{IJBC}),
these
are also found to persist when the coupling is switched on.
Over the years, some of the DB structures have been
endowed with distinguishing names. The Sievers-Takeno mode corresponds to a
DB with only one excited site (it can also be denoted as single site or 1-site breather). Also,
the Page mode is a multibreather where two adjacent sites are excited; notice that the former is a site-centered breather whereas the latter corresponds to a bond-centered (inter-site-centered) one. When all the sites at the AC limit are excited, we are dealing with a nonlinear phonon or {\em phonobreather}; a dark breather \cite{Dark} is a phonobreather with one (or a few) non-excited site(s),
resembling the functional form of a dark soliton of the nonlinear
Schr{\"o}dinger equation.

Multibreathers are usually formed by time-reversible oscillators. In these cases, they can be characterized by a coding sequence $\sigma\equiv\{\sigma_n\}$ indicating the phase and the excitation state of the relevant sites at the AC limit. This code is $\sigma_n=0$ if the $n$-th oscillator is at rest, $\sigma_n=1$ if it oscillates with a frequency $\wb$ and phase $0$ (i.e. $u_n(0)>0$) and $\sigma_n=-1$ if oscillates with initial phase $\pi$ (i.e. $u_n(0)<0$). For instance, the Sievers-Takeno and Page modes are represented by $\sigma=\{1\}$ and $\sigma=\{1,1\}$ (for soft potentials) or $\sigma=\{1,-1\}$ (for  hard potentials), respectively. There are some cases where the code can be more complex as in sine-Gordon potentials, where rotors can coexist with oscillators; in such cases, structures called {\em rotobreathers} (see \cite{Aubry,Takeno}) can emerge if the excited site corresponds to a rotor; they have been experimentally generated in Josephson junction arrays \cite{Trias,Ustinov}.

There are also, however, some special cases where multibreathers are constituted by non-time-reversible oscillators, as demonstrated in \cite{Aubry}. One can find such kind of solutions in 1D chains with periodic boundary conditions in the form of phonobreathers with phase torsion \cite{Cretegny2,ourphonon}, or in 2D percolating clusters or so-called discrete
vortices \cite{Cretegny1,Vassilis1,IJBC}. They can also emerge in 1D lattices with long-range interactions \cite{Vassilis2,Rapti}. Additionally, they
constitute a potential attractor
in periodically forced and damped oscillator networks~\cite{Marin2}.

The non-resonance condition of MacKay-Aubry's theorem clearly establishes
that the breather frequency $\wb$ must lie in the gaps between the linear modes (phonons) band; additionally harmonics of this frequency must avoid
resonances with the band to ensure the absence of
energy dispersing mechanisms affecting the DB.
In the case of 1D KG lattices, there is an optical band of phonons given by
\begin{equation}\label{eq:phonon}
    \wp^2=\wo^2+4C\sin^2\frac{q}{2}
\end{equation}
where $q$ is the phonon wavenumber and $\wp$ the frequency of the
associated effectively plane wave excitations $\sim e^{i (q n-\wp t)}$.
In finite lattices (which are needed for numerical computations), the value of $q$ is quantized with the quantization being determined by the
nature of the imposed boundary conditions and the number of lattice nodes $N$. Consequently, breathers in hard potentials, can be continued  until $\wb$ collides with the upper edge of the phonon band ($q\approx\pi$), an event occurring when $C\approx(\wb^2-\wo^2)/4$ (the approximation symbol is used to
take into account the fact that, depending on the quantization of $q$, it could happen that $q=\pi$ is not in the band). Resonances in soft potentials are caused by integer multiples of the breather frequency colliding with the
frequencies of the phonon band. The relevant critical point emerges
when the second harmonic in asymmetric potentials and the third one in symmetric ones (like the purely quartic $\phi^4$ analyzed in this Chapter) collides
once again with the upper edge of the phonon band; that is, when $C\approx\wb^2-\wo^2/4$ or $C\approx(9\wb^2-\wo^2)/4$, respectively. In addition, in a finite lattice and soft potentials, there are gaps in the phonon band and breathers can ``bypass'' the resonance frequency, by existing within these finite
gaps. In this case, the breather hybridizes with the bifurcating phonon creating a structure called \emph{phantom breathers}; strictly speaking, they are non-exponentially-localized breathers as they have a tail oscillating  with a frequency $n\wb$ with $n$ depending on the multiple of the breather frequency that resonates \cite{phantom}. These structures are the
discrete analogue of the nanoptera observed at the continuum limit of KG lattices with the $\phi^4$ double well potential~\cite{Boyd}. As shown in \cite{Martina} for sine-Gordon and $\phi^4$ double well potentials, continuation up to the continuum limit is similar to a Wannier-Stark ladder. This phenomenon (i.e. breather-phonon hybridization) takes place because the staggering character of the phonons is different
than that of the breather; if the breather tails had the same staggering character of the bifurcating phonon, the breather would be smoothly continued from the phonon and its amplitude would be zero at the phonon frequency. Notice that gaps in phonon band appears naturally in disordered lattices because of Anderson localization; breathers in such systems can delocalize \cite{Kopidakis1} or remain localized \cite{Kopidakis2,Archilla}.
The above scenario is typical of Sievers-Takeno and Page modes. When other multibreathers are considered, breathers cannot reach the phonon band and the bifurcation scenario is more complex (see e.g. \cite{Martina}).

Let us also mention that decay of breathers is not necessarily
exponential. In lattices with long-range intersite interactions, the decay
can be algebraic \cite{Flach3}, possibly featuring a transition
from exponential to algebraic; for a recent experimental
realization of this transition in system based on magnets,
see, e.g.,~\cite{cho}. In addition, in some KG/FPUT lattices with
$\phi^4$ intersite potential, breathers can decay superexponentially being
dubbed as (nearly) compact DBs \cite{Dey,Comte}.

Finally, we must remark that there are several more existence proofs. For instance, a variational proof was introduced in \cite{Vesna}; this is valid only for hard potentials. A proof based on the center manifold theorem was introduced in \cite{Guillaume} and is valid for DBs whose frequency is close to the phonon
band edge.

\section{Stability of discrete breathers}

This Section can be considered as the core of the present Chapter. The treatment of the topic will be as follows: first of all, we will present an introduction to the Floquet theory applied to the linear stability of discrete breathers; then, we will show different approaches to the linear stability of multibreathers in the vicinity of the AC limit and introduce the energy-vs-frequency
monotonicity criterion for the linear stability of breathers and multibreathers at arbitrary coupling; afterwards, nonlinear stability criteria will be
summarized together with the dynamical evolution of some examples of unstable solutions.

\subsection{Floquet analysis}

In order to assess the dynamical robustness of the identified
DB solutions and their potential accessibility in physical experiments,
a key step is the
determination of their stability. In the paper where MacKay and Aubry demonstrate the existence of breathers \cite{MA}, it is speculated that 1-site breathers are likely to be stable. This was finally proven by Aubry in~\cite{Aubry} for finite lattices and MacKay and Sepulchre in~\cite{Sepulchre} for infinite
lattices.

As breathers are time-periodic solutions of the equation system (\ref{eq:dyn0}), their spectral stability can be determined by means of a Floquet analysis. To this aim, we need to evaluate the evolution of a perturbation $\xi_n(t)$ to a solution $v_n(t)$. We thus introduce ---in e.g. the 1D equation (\ref{eq:dyn1d})--- the solution $u_n(t)=v_n(t)+\epsilon\xi_n(t)$ with $\epsilon$ being a small constant. Then, the equation that the perturbation satisfies
to $\mathcal{O}(\epsilon)$ is

\begin{equation}\label{eq:perturb}
    \ddot{\xi}_n+V''(v_n)\xi_n+C(2\xi_n-\xi_{n+1}-\xi_{n-1})=0.
\end{equation}

This equation can be written in a more compact form as

\begin{equation}\label{eq:Newtonop}
    \mathcal{N}(v(t))\xi=0
\end{equation}
where $\xi\equiv\{\xi_n(t)\}$ and $v(t)\equiv\{v_n(t)\}$. $\mathcal{N}$ is known as the linearization operator. If $\xi\in\mathcal{C}^2$, the study of
this operator spectrum offers information about stability. If, moreover, $\xi\in\mathcal{E}_s^2(\wb)$ (i.e. $\xi$ belongs to the space of time-reversible solutions with frequency $\wb$), this operator can be identified
as the Jacobian (Fr\'echet derivative) of the dynamical equations, i.e. $\partial_vF(v(t))=\mathcal{N}(v(t))$. Equation (\ref{eq:Newtonop}) can be viewed as a the particular case $E=0$ of the eigenvalues equation for the Newton operator

\begin{equation}\label{eq:Newtonop2}
    \mathcal{N}(v(t))\xi=E\xi
\end{equation}

The study of the spectrum of $\mathcal{N}$ is related to Aubry's band theory \cite{Aubry} as we will see further in the present section. In Hamiltonian dynamical systems, this linearization operator is time-symmetric, real, symplectic and Hermitian. In addition, it is invariant on time translations of period $T$, so by virtue of Bloch's theorem, the corresponding
eigenfunctions $\xi(t)$ can be expressed as Bloch functions:

\begin{equation}\label{eq:Bloch}
    \xi(t)=\e^{i\theta t/T}\upsilon(t).
\end{equation}

To perform Floquet analysis, we need to study the spectrum of the Floquet operator $\mathcal{F}$, defined from the following map:

\begin{equation}\label{eq:Floquet}
    \Omega(T)=\mathcal{F}_o\Omega(0),\qquad \mathrm{with}\ \Omega(t)=[\xi(t),\dot{\xi}(t)]
\end{equation}

The representation of $\mathcal{F}_o$ in $\mathbb{R}^{2N}$ is denoted as the
so-called monodromy matrix. The eigenfunctions of $\mathcal{F}_o$ are those of the Newton operator with $E=0$. Thus, as a consequence of Bloch's theorem (\ref{eq:Bloch}), $\Omega(T)=\exp(i\theta)\Omega(0)$. In other words, monodromy eigenvalues (also known as Floquet multipliers) are of the form $\lambda=\exp(i\theta)$ with $\theta\in\mathbb{C}$. $\theta$ is dubbed as Floquet argument.

As the linearization operator is real and symplectic, the Floquet operator also possesses both properties. From the fact that $\mathcal{F}$ is real, one can deduce that if $\lambda$ is a Floquet multiplier, $\lambda^*$ is also a multiplier. Because of the symplecticity of $\mathcal{F}$, $1/\lambda$ is also a multiplier. In other words, Floquet multipliers always come in quadruplets $(\lambda,\lambda^*,1/\lambda,1/\lambda^*)$ if $\lambda\notin\mathbb{R}$ and in pairs $(\lambda,1/\lambda)$ if $\lambda\in\mathbb{R}$. Consequently, a necessary and sufficient condition for a Hamiltonian dynamical system to be linearly stable is that $\theta\in\mathbb{R}$ (i.e. that the multipliers lie at the unit circle of the complex plane). If the system is not Hamiltonian, the symplecticity can be broken and the previous property does not hold. Then, the condition for stability is that $|\lambda|\leq1$.

An important concept in Floquet analysis is the Krein signature, which, for a given eigenvalue $\lambda$ is defined as
\begin{equation}\label{eq:Krein}
    \kappa(\lambda)=\mathrm{sgn}\left(\sum_{n=1}^N\mathrm{Im}\left[\xi_n^*(t)\dot{\xi}_n(t)\right]\right)
\end{equation}
Given the symplectic nature of $\mathcal{N}$, the total Krein signature (i.e. the sum of the Krein signature of every Floquet multipliers) is conserved in time. If $\lambda$ is real, $\xi_n\in\mathbb{R}\ \forall n$ and $\kappa(\lambda)=0$. The Krein signature is helpful for predicting bifurcations when system parameters vary: due to the properties of Floquet multipliers in Hamiltonian systems, the only way that an instability takes place is that a pair of eigenvalues coincide at $\theta=0$ or $\theta=\pi$ or a quartet of eigenvalues coincide at a $\theta$ different than $0$ or $\pi$. However, the coincidence of eigenvalues is not a necessary condition for an instability to take place. Krein's criterion establishes that a necessary condition for the instability emergence is that the coincident multipliers have different Krein signature. Nevertheless, this criterion is not a sufficient condition as discussed, e.g., in the context of Aubry's band theory in~\cite{Aubry}.

The starting point of the band theory is Eq. (\ref{eq:Newtonop2}), from which Bloch's theorem establishes that eigenvalues come in a set of non-overlapping bands $E_\nu(\theta)$, which is a continuous function, non constant and $2\pi$-periodic. From (\ref{eq:Bloch}), one can write $\xi(T)=\exp(i\theta_\nu(E))\xi(0)$. $\theta_\nu$ can be chosen in the first Brillouin zone, i.e. $\theta_\nu\in(-\pi,\pi]$. Then, the set of eigenvalues with $\theta$ in the first Brillouin zone is denoted as band $\nu$. Bands are asocciated with stable solutions if $\theta_\nu\in\mathbb{R}$. Moreover, due to the properties of the linearization operator, bands are symmetric with respect to $\theta_\nu=0$. The values $\theta_\nu(E)$ are found by diagonalizing matrix $\mathcal{F}_E$, which is found by applying (\ref{eq:Floquet}) to the integration of (\ref{eq:Newtonop2}). The monodromy coincides with $\mathcal{F}_0$ and, consequently, the Floquet arguments are $\theta_\nu(0)$. That is, in order for the solution to be stable, there must exist $2N$ bands that cut or are tangent to the $E=0$-axis. As mentioned above, the band analysis allows to improve Krein's criterion: as demonstrated in \cite{Aubry}, the Krein signature of a Floquet multiplier is minus the sign of the slope at $E=0$ of the band corresponding to this multiplier; i.e.

\begin{equation}
    \kappa(\theta_\nu)=-\mathrm{sgn}\left(\frac{\mathrm{d}E_\nu(\theta)}{\mathrm{d}\theta}\big{|}_{\theta=\theta_\nu}\right)
\end{equation}

As discussed in \cite[Section 4.4]{Aubry} a necessary and sufficient condition for a bifurcation to occur is that the coincident eigenvalues belong to the same band, as in that case the slopes at the bifurcation point will have opposite signs.

Floquet analysis can also help to identify what kind of bifurcations a breather experiences. If a pair of eigenvalues collides at $\theta=0$, the breather undergoes an exponential (tangent) bifurcation, whereas collisions at $\theta=\pi$ correspond to a period-doubling bifurcation. In the rest of cases, there is a quartet of multipliers colliding out of the real axis and the breather experiences a (Hamiltonian) Hopf bifurcation \footnote{Notice that, strictly speaking, we are dealing with a Neimark-Sacker bifurcation (Hopf bifurcation of periodic orbits). However,  through a slight abuse of the relevant terminology, we denote it simply as a Hopf bifurcation.}.

Prior to show the stability properties of breathers, we want to indicate that when the dynamical equations are invariant under time translation, there is a pair of eigenvalues that always remain at $\theta=0$, known as the phase and growth mode, which come from the fact that $\xi_n=\dot{u}_n$ and $\xi_n=\partial u_n/\partial \wb$ are solutions of Eq. (\ref{eq:perturb}). The growth mode is, in
fact, a marginal mode, as a perturbations along its direction growth linearly.
This mode is associated to a generalized eigenvector along this
eigendirection. These modes will be particularly important when considering
stability criteria below.

\subsection{Linear stability near the anti-continuous limit}
\label{sec:stabAC}

Prior to introducing the existing approaches for the stability in the vicinity of the AC limit, we need to understand the structure of the monodromy spectrum at that limit. Let us suppose that the lattice is formed by $N$ oscillators and we have a multibreather with $p$ excited sites at $C=0$. Then, it is easy to deduce from (\ref{eq:perturb}) that there are $p$ pairs of degenerated phase/growth modes at $\theta=0$ corresponding to excited sites, and $N-p$ pairs of multipliers at $\theta=\pm\wo T\ \mathrm{mod}\ 2\pi$ corresponding to the oscillators at rest. The Krein signature of the multipliers at $\theta=0$ is zero, but for the oscillators at rest, it depends on the hardness/softness of the substrate potential, the breather frequency and the sign of the coupling constant. From now on, let us suppose that $C>0$. In that case, the multiplier in the upper half-circle will have $\kappa=1$ (and those of the lower half-circle will have $\kappa=-1$) when $\wb>2\wo$ if the on-site potential is hard and $\frac{2}{2k+1}\wo<\wb<\frac{1}{k}\wo$ with $k\in\mathbb{N}$ if the potential is soft. On the contrary, if $\wo<\wb<2\wo$ for hard potentials and $\frac{1}{k+1}\wo<\wb<\frac{2}{2k+1}\wo$ for soft ones, the multipliers at the upper half-circle will have $\kappa=-1$ \cite{Sepulchre,Floria}.

When the coupling is switched on, we need to analyze the fate of the multipliers at $\theta\neq0$, that will correspond to the linear modes, and those at $\theta=0$. The former will expand so that the isolated eigenvalue pair at $\theta\neq0$ become a couple of arcs with $\theta\approx\pm\wp T\ \mathrm{mod}\ 2\pi$, being $\wp$ the phonon frequencies (\ref{eq:phonon}). Notice that the Krein signature of the eigenvalues of an arc is the same as the multiplier from which they are born.
With this in mind, it is easy to deduce that a 1-site breather must remain stable when the coupling is switched on because, on the one hand, there are no bifurcations stemming from the linear mode arcs and, on the other hand, $p=1$ and the only pair at $\theta=0$ corresponds to the phase/growth modes and must remain there for every coupling. Figure \ref{fig:example} shows a couple of examples of Floquet multipliers spectrum for a 1-site breather in soft and hard potentials.

When dealing with a multibreather, $p>1$ and there are $p-1$ eigenvalues whose evolution is not clear (i.e. if they move along the circle or they abandon it through the real axis). The first attempt to give a generic explanation
of the emerging possibilities
was based on Aubry's band theory \cite{ourbands}, while
a second approach, which makes use of the effective Hamiltonian
methodology~\cite{VassilisPanos}, is able to prove the existence of breathers at small coupling and also to give an expression of the
individual multipliers at low coupling. The work of \cite{IJBC} demonstrated
that the two approaches are equivalent if one assumes that Aubry's bands are parabolic. The main drawback of these findings is that they can only predict the stability or instability of multibreathers at low coupling when all the excited sites are adjacent. The latter limitation
was overcome in the work of \cite{Sakovich} by considering higher order
perturbations. However, for simplicity, we will hereafter
focus on the description of multibreathers whose excited sites are adjacent,
as per the earlier studies by the authors in~\cite{IJBC,ourbands,VassilisPanos}.

As a general result, we proved that, for $C\rightarrow0^+$, multibreathers are stable when all the excited sites at the AC limit are in phase (anti-phase) if the potential is hard (soft). Two sites $i$ and $j$ are said to be in phase (anti-phase) if $\sigma_i\sigma_j=+1$ ($\sigma_i\sigma_j=-1$). This property is reversed if $C<0$; such a case is relevant when considering e.g. dipole-dipole interactions in DNA models \cite{ourdna}. If the excitation pattern is neither
uniformly in phase, nor uniformly in anti-phase (e.g. $\sigma=\{1,-1,-1,1\}$) the multibreather is unstable, independently of the hardness of the potential or sign of coupling constant, as demonstrated in \cite{ourbands2}. The work
of~\cite{VassilisPanos} provided a detailed count of the
unstable eigendirections in such a case and an estimate of the
associated multipliers in the small $C$ limit.

As an instructive special case example, let us focus on
the simplest multibreather case, namely the 2-site DB in the hard and soft $\phi^4$ potential (\ref{eq:phi4potential}). Among these multibreathers, the ones that are unstable close to the AC limit, by virtue of the theorems of \cite{ourbands,VassilisPanos}, are the Page modes whose codes are $\sigma=\{1,1\}$ (if the potential is soft) and $\sigma=\{1,-1\}$ (if the potential is hard). The theorems also predict that one of the multiplier pairs lying at $\theta=0$ at the AC limit moves along the real axis with its (imaginary) argument given by:

\begin{equation}\label{eq:arg1}
    \theta=\pm \frac{2i\pi}{\wb}\sqrt{2sC\frac{J}{\wb}\frac{\partial \wb}{\partial J}},
\end{equation}
with
\begin{equation}\label{eq:arg2}
    J=\frac{1}{2\pi}\int_0^T[\dot u(t)]^2\mathrm{d}t=\frac{\wb}{2}\sum_{k\geq1} k^2z_k^2,
\end{equation}
being the action of an isolated oscillator. In the case of
the $\phi^4$ potential, finding an analytical form of $\partial \wb/\partial J$ by using (\ref{eq:1oschard}) and (\ref{eq:1oscsoft}) is a cumbersome task. Because of this, it is easier to calculate the Fourier coefficients in (\ref{eq:arg2}). An alternative is to use the rotating wave approximation (RWA) $u(t)\approx z_1\cos(\omega t)$, obtaining that:

\begin{equation}
    z_1=2\sqrt{\frac{\wb^2-1}{3s}}
\end{equation}

\begin{figure}[tbp]
\begin{center}
\begin{tabular}{cc}
\includegraphics[height=4.5cm]{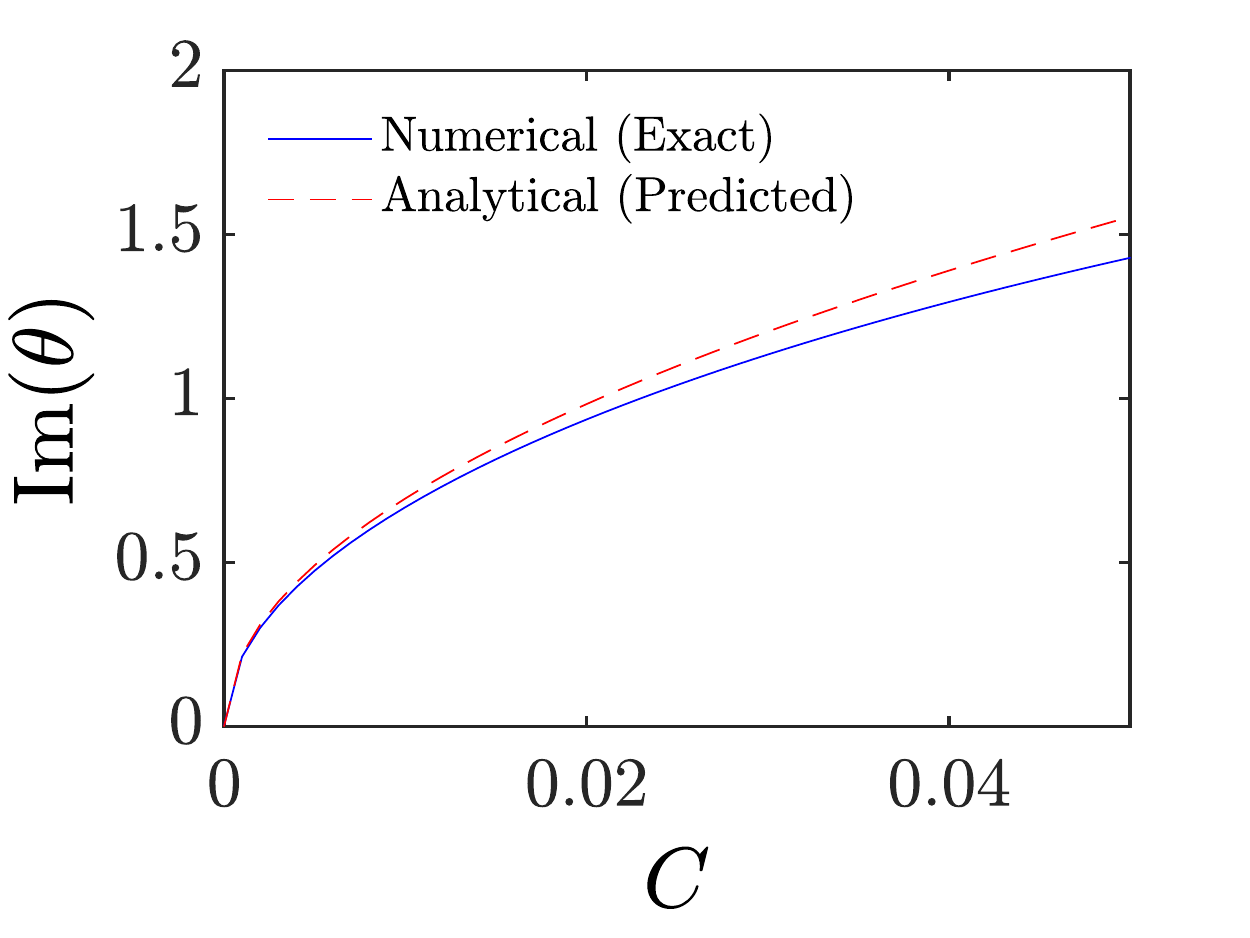} &
\includegraphics[height=4.5cm]{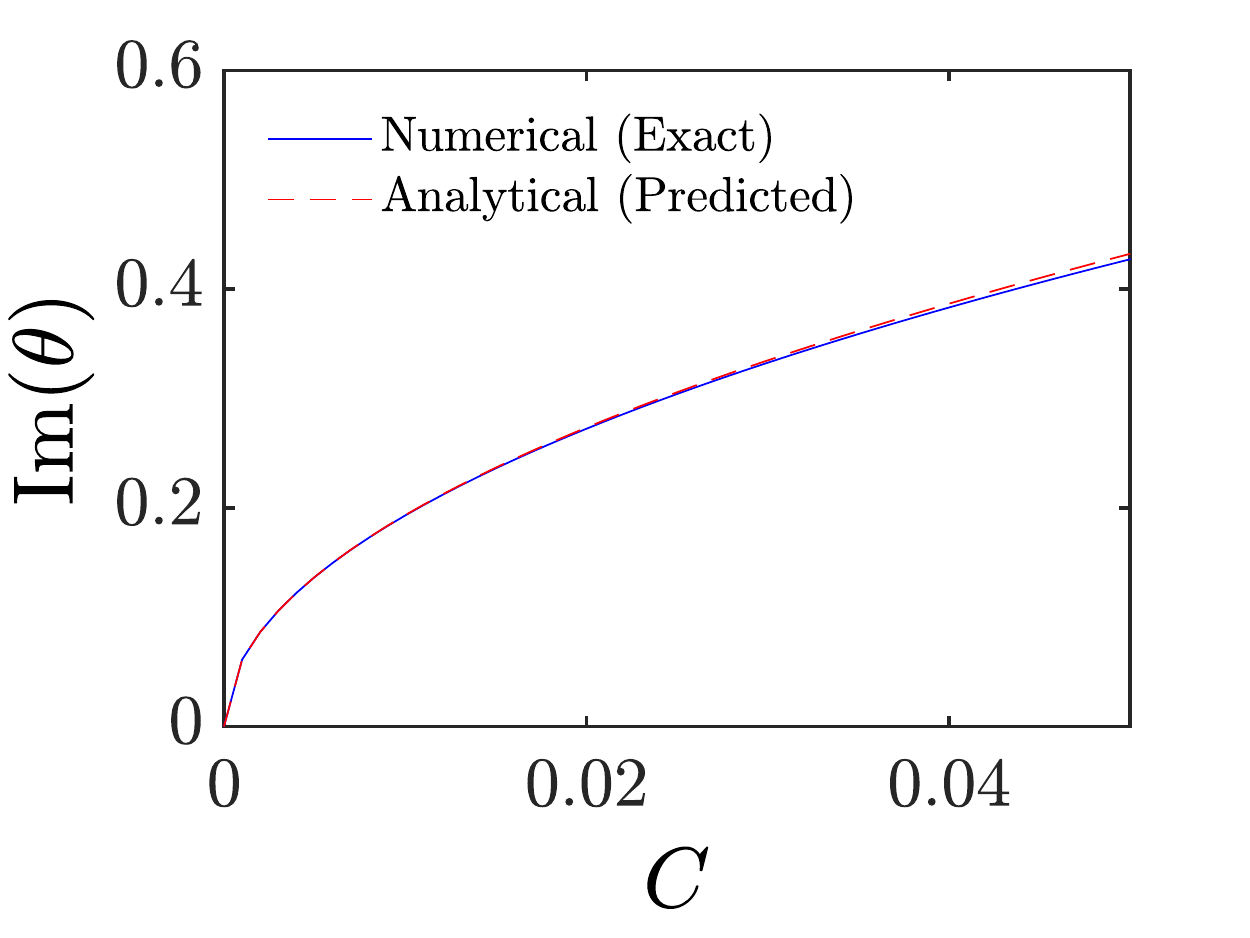} \\
\end{tabular}
\end{center}
\caption{Imaginary part of the argument of the Floquet multiplier responsible for the instability of Page modes with respect to $C$ in soft potentials with $\omega_\mathrm{b}=0.80$ (left panel) and hard potentials with $\omega_\mathrm{b}=2.5$ (right panel). The figure, which highlights the region of small coupling, also shows the  analytical prediction of Eq. (\ref{eq:argrwa}).}
\label{fig:FloqAC}
\end{figure}

Then, the Floquet argument of the analyzed multibreathers can be written as:
\begin{equation}\label{eq:argrwa}
    \theta=\pm \frac{2i\pi}{\wb}\sqrt{2sC\frac{\wb^2-1}{3\wb^2-1}}.
\end{equation}
In Figure~\ref{fig:FloqAC}, we compare the Floquet arguments of the stable 2-site multibreathers found by integrating the perturbation equation (\ref{eq:perturb}) with those of the predictions of \cite{IJBC,VassilisPanos}, i.e. Eq. (\ref{eq:argrwa}) for $C\rightarrow0^+$. Notice the good agreement between all the curves.

The stability theorems have also been useful towards investigate the stability or instability (at low coupling) of dark breathers \cite{ourbands} and multibreathers in the presence of inhomogeneities \cite{ourbands2}. They have also been applied to multibreathers with broken time-reversibility, as phonobreathers with phase torsion \cite{ourphonon}, multibreathers in 1D lattices with long-range interactions \cite{Vassilis2,Rapti} and vortices in triangular \cite{Vassilis1} and rectangular latices \cite{IJBC}. In the latter case, the stability theorems are unable to get analytical predictions when the on-site potential is even (as e.g. in the hard and soft $\phi^4$ potentials), as the perturbations are of order higher than $\sim\sqrt{C}$. This type of scenario has been explored
in the case of the simpler class of discrete nonlinear
Schr{\"o}dinger models, as summarized in~\cite{dnls}. In that context,
it has been dubbed as the super-symmetric case, and requires
higher order perturbative analysis in order to identify the leading order
contributions to the multipliers for which the corresponding discrete
vortices arise at O$(C)$ --- although for some of them, they arise at much
higher order (such as O$(C^3)$). A systematic consideration of
such stability findings for DBs is a worthwhile investigation for
future work. It should be noted that in the present contribution
we do not expand further on the specifics of both the Aubry band
and the effective Hamiltonian methods and their analytical
predictions and comparisons with full numerical results. A reasonably
recent and relatively
up-to-date summary of these methods in more detail can be found
in~\cite{IJBC}.

\subsection{Linear stability at arbitrary coupling. An energy-based stability criterion}

When the coupling is increased from the situation described in the previous Subsection, the phonon arcs will be expanding. The mapping between angles and phonon frequencies is not exact because there are hybridization between the linear modes and the breather itself, which is more pronounced when the coupling is
progressively higher. In fact, in many cases, beyond a threshold coupling, exponentially localized so-called ``internal modes''
can bifurcate from the phonon band. These modes localize in the rightmost (leftmost) edge of the arc if the Krein signature of the phonons of the upper half-circle is positive (negative). In addition, in the case of multibreathers, there can also exist localized multipliers that were at $\theta=0$ at the AC limit.
These modes may move on the unit circle for finite coupling and may lead
to various scenarios of collision between Floquet multipliers. For instance,
two pairs of localized modes can collide at $\theta\neq0$ or a localized mode
pair can collide with the phonon arcs. Recall that the Krein signature of the colliding multipliers must be opposite in order for a (Hamiltonian Hopf)
bifurcation to take place
(otherwise the pairs are led to a so-called avoided crossing scenario
without producing an instability). In the opposite Krein multiplier
collision case, we are dealing with a Hopf bifurcation (also called Krein crunch) that, in most cases, persists when the lattice is enlarged. Another situation occurs when the phonon arcs themselves collide. As they have opposite Krein signature, more Krein crunches can take place. If this bifurcation takes place at $\theta=0$, it is a signal that a resonance of the breather with the phonon band has taken place and, consequently, the existence condition of the MacKay-Aubry theorem has been violated. If the arcs' overlap takes place at $\theta\neq0$, the bifurcations are generally ruled out at infinite lattices~\cite{finite}.

There are other interesting cases. For instance, when localized modes collide
at $\theta=\pi$, they bring about a period-doubling bifurcation; such a bifurcation was found for $\phi^4$ Hamiltonian lattices at low coupling in \cite{Sakovich}. Another scenario corresponds to the case when a multiplier, which escapes from the phonon arc, corresponds to an anti-symmetric mode and moves along the unit circle, eventually colliding with its conjugate at $\theta=0$; in this case, the mode is structurally similar to the translational one and a perturbation of a breather along its direction, which increases linearly as it is a marginal mode, gives rise to a moving breather \cite{MovingCretegny}. We will deal with this topic in more detail in Subsection \ref{sec:moving}.

\begin{figure}[tbp]
\begin{center}
\begin{tabular}{cc}
\includegraphics[height=4.5cm]{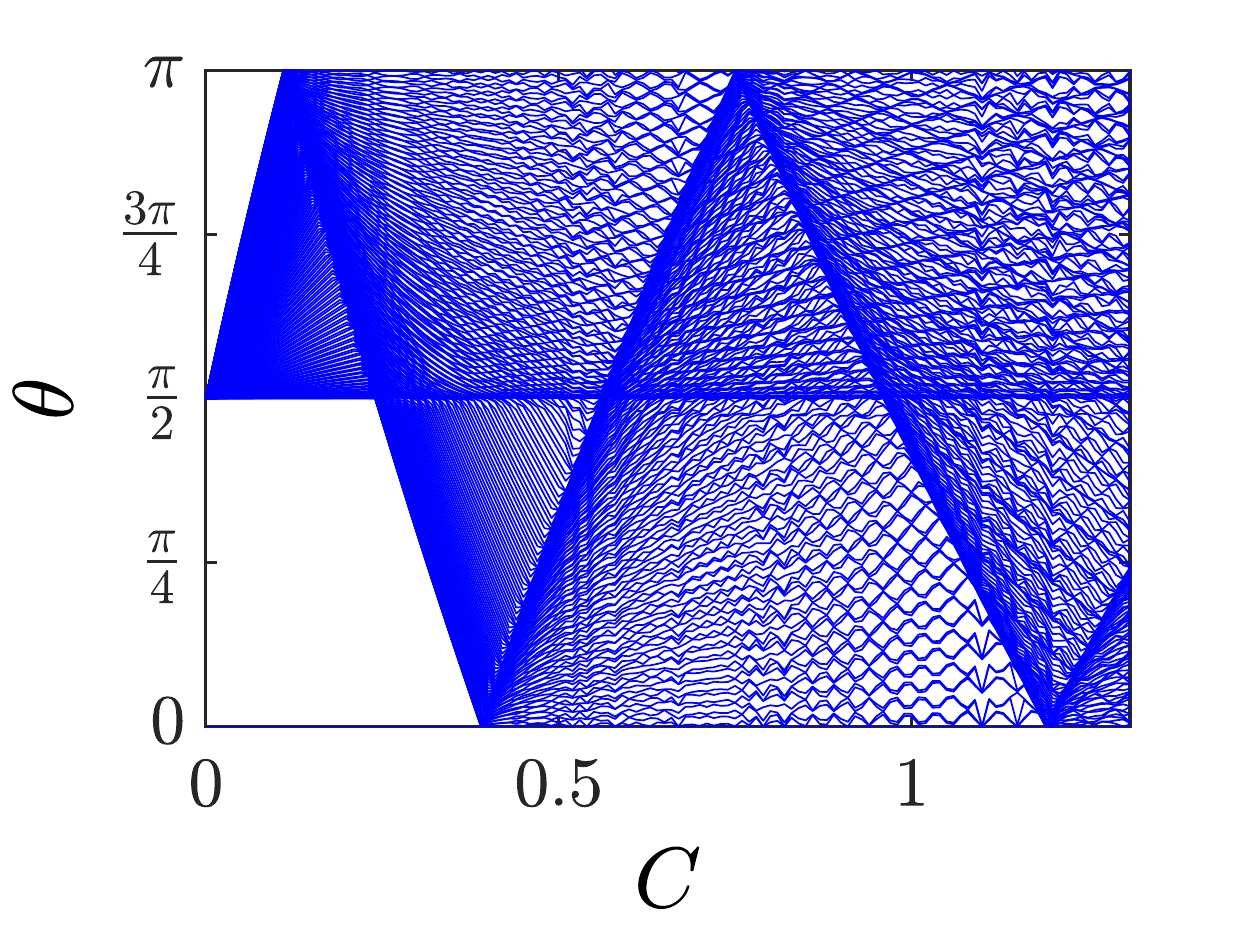} &
\includegraphics[height=4.5cm]{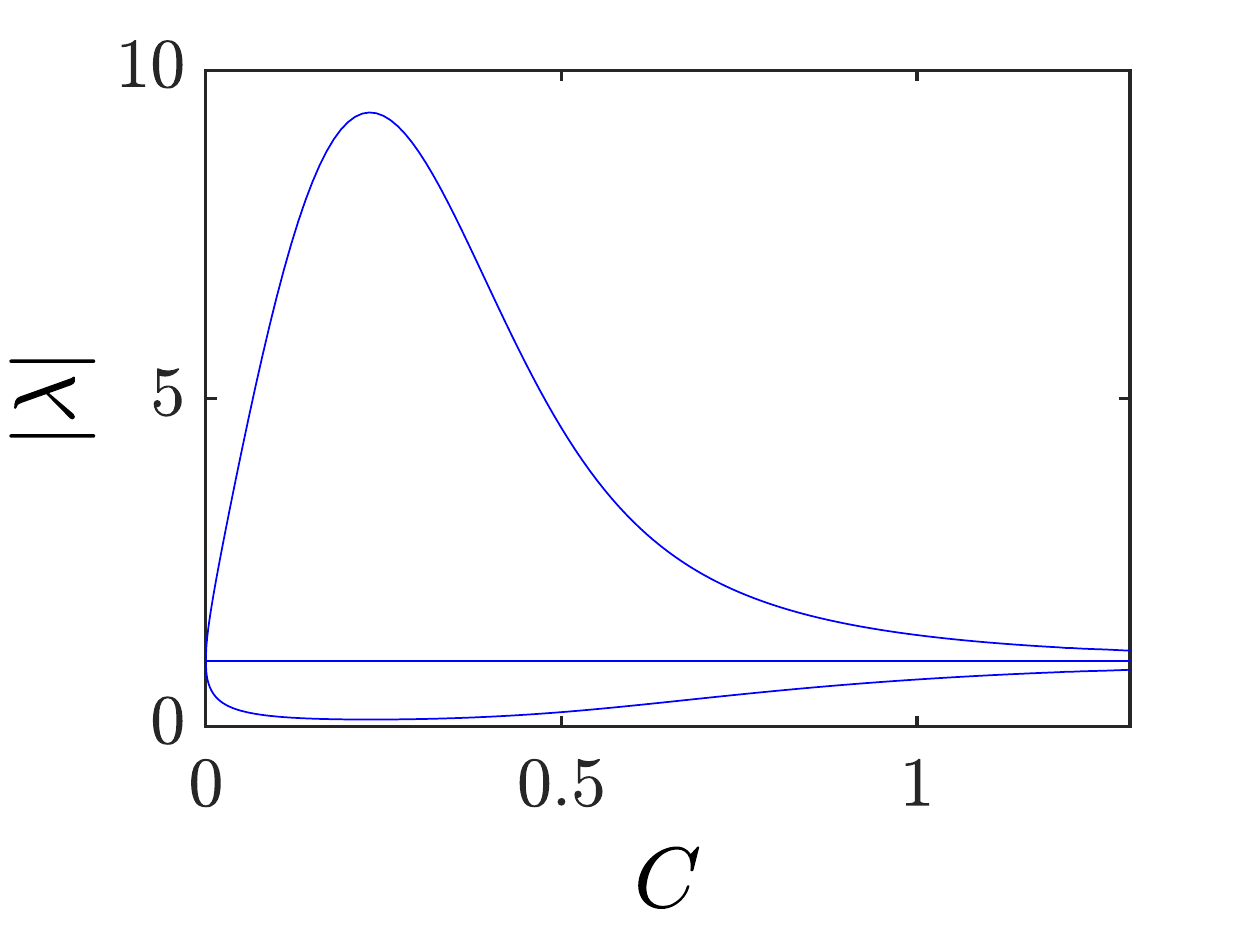} \\[3mm]
\includegraphics[height=4.5cm]{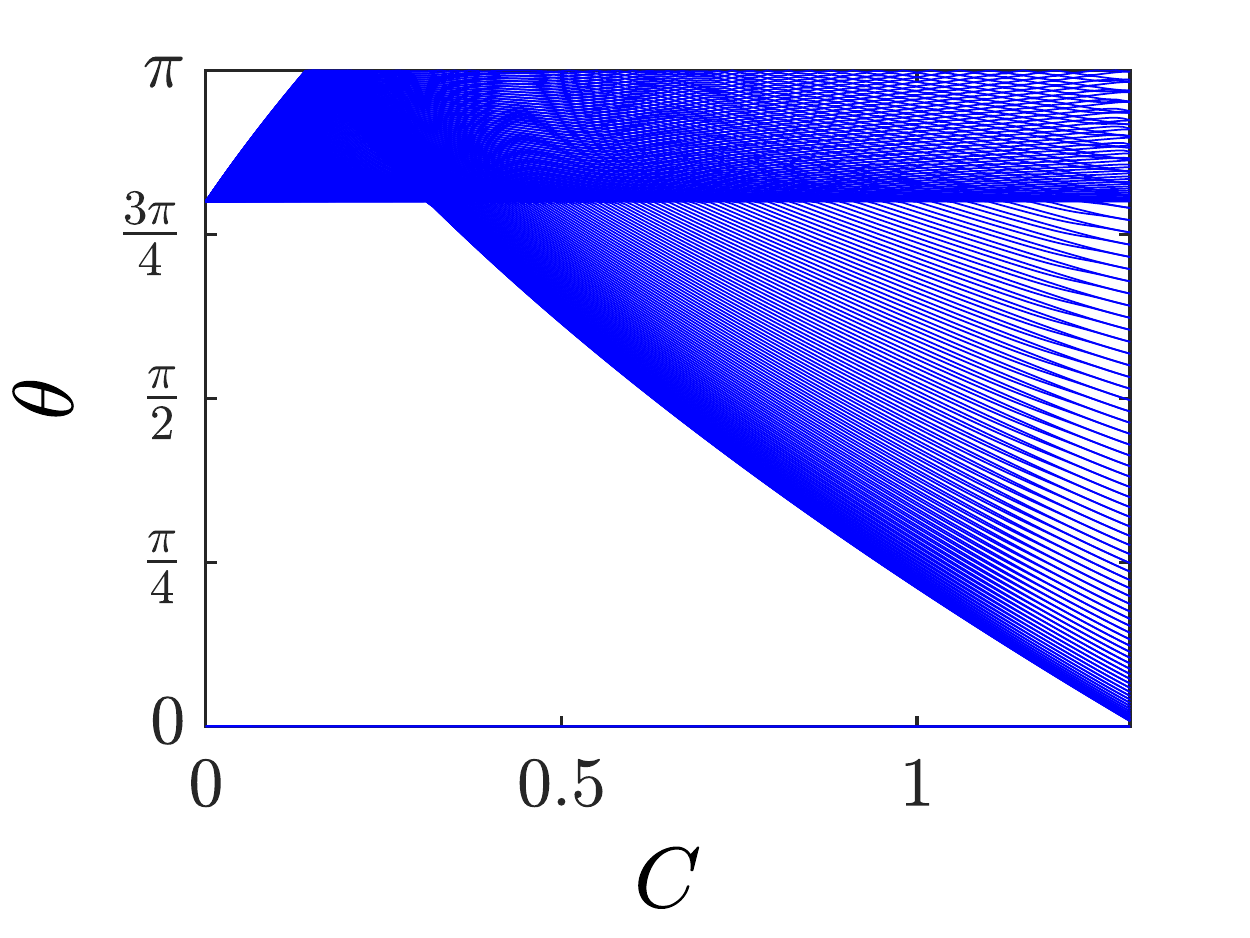} &
\includegraphics[height=4.5cm]{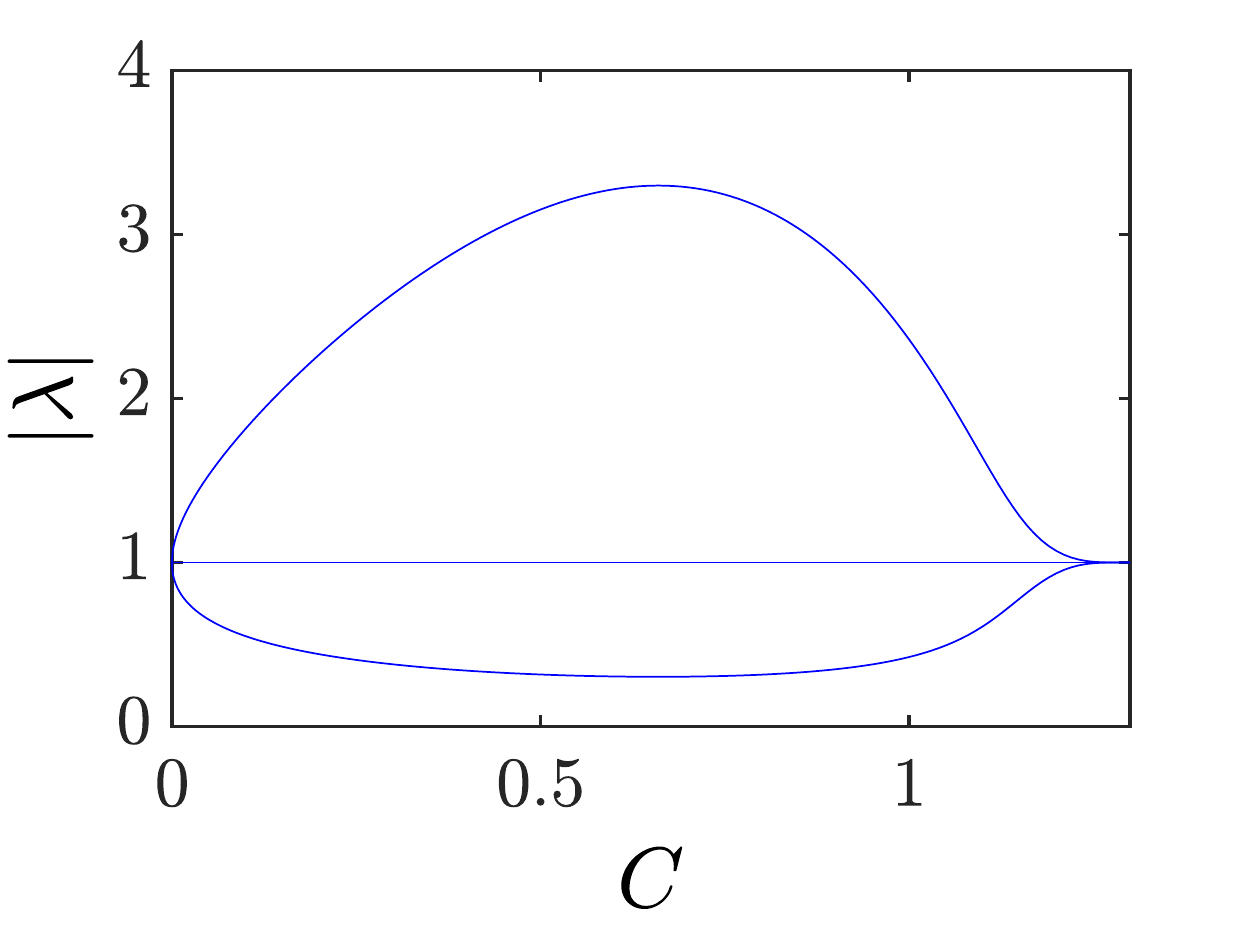} \\[3mm]
\includegraphics[height=4.5cm]{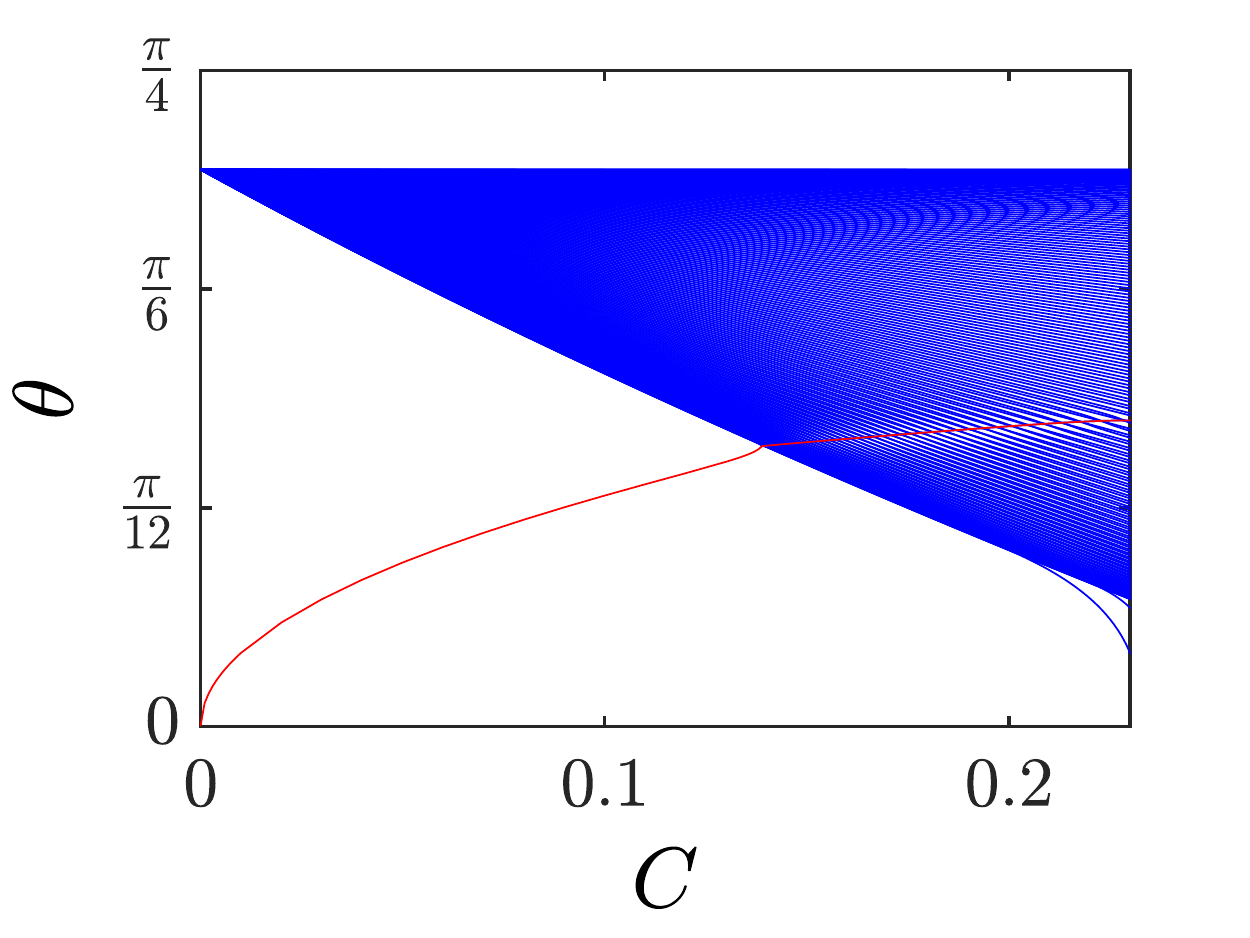} &
\includegraphics[height=4.5cm]{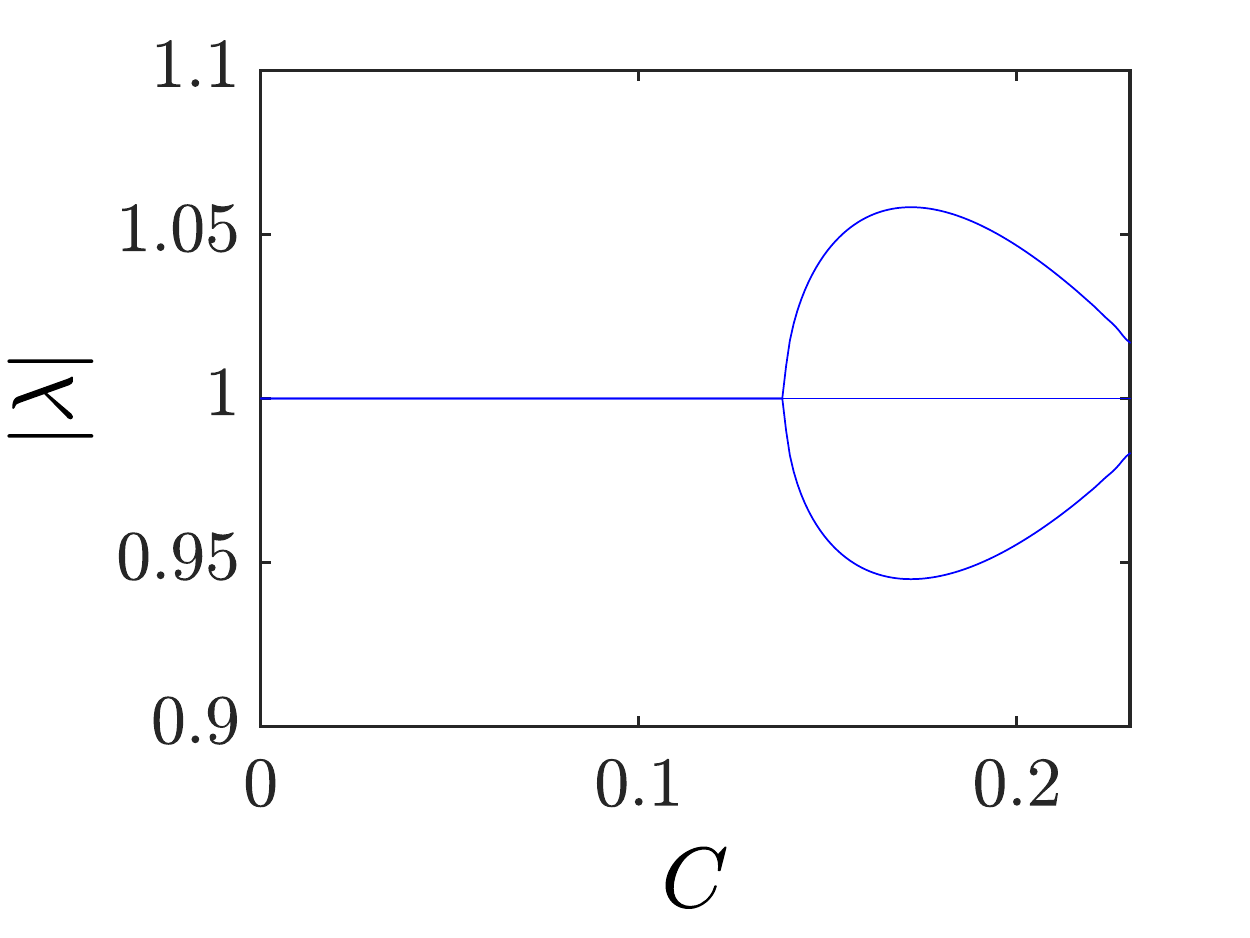} \\
\end{tabular}
\end{center}
\caption{Dependence with respect to the coupling constant $C$ of the argument (left panels) and modulus (right panels) of the Floquet multipliers corresponding to the following configurations: (top panels) $\sigma=\{1,1\}$ multibreather with $\omega_\mathrm{b}=0.8$ in soft potentials; (middle panels) $\sigma=\{1,-1\}$ multibreather with $\omega_\mathrm{b}=2.5$ in hard potentials; (bottom panels) $\sigma=\{1,1\}$ multibreather with $\omega_\mathrm{b}=1.5$ in hard potentials. Notice that in the right panels, the mode with positive (negative) Krein signature is depicted in red (blue).}
\label{fig:stability}
\end{figure}

Figure \ref{fig:stability} shows examples of the Floquet multiplier dependence with respect to $C$ for Page (unstable two-site) modes
and also for a stable configuration close to the AC limit, which corresponds to the $\sigma=\{1,1\}$ multibreather in the hard potential. In the case of
the Page modes one can see that the only instability is the one caused by the Floquet multiplier that abandons the circle at the AC limit, and that, in the case of hard potentials, this mode comes back to the circle when the breather frequency resonates with the linear modes band.
For the $\sigma=\{1,1\}$ multibreather in the hard potential, it is observed that the instability is caused by the collision with the phonon arc of the localized mode departing from $\theta=0$ at the AC limit. Notice that the $\sigma=\{1,-1\}$ multibreather, which is stable for small coupling when the on-site potential is soft, undergoes a Hopf bifurcation caused by the collision of two localized modes in KG lattices with Morse on-site potential (see e.g. \cite{SAM}); however, in soft $\phi^4$ potentials, this bifurcation does not take place and, instead, there is a Hopf bifucation similar to the one of the $\sigma=\{1,1\}$ multibreather in the hard potential.

Tangent bifurcations can also occur when there exists an extremum in the energy dependence with respect to the breather frequency. In Klein-Gordon lattices, such extrema seem only to appear in two-dimensional lattice and, in several cases, are related to energy thresholds for the existence of breathers \cite{Mica,Kladko,Kastner}. The stability criterion that we analyzed in \cite{Energy} is related to such bifurcations. This criterion is a generalization to Klein-Gordon and FPUT lattices (and their DB solutions)
of the well known Vakhitov-Kolokolov (VK) criterion defined for the Nonlinear Schr\"odinger (NLS) Equation \cite{Vakhitov}. In fact, as shown in~\cite{Energy},
upon the multiscale expansions that lead from the former equations
to the latter one, the VK criterion is retrieved as a special case
example.
This establishes a sufficient condition for the instability of breathers. In Klein-Gordon lattices, a family of breathers is exponentially unstable if its energy increases with the frequency, if the on-site potential is soft, or if the energy decreases with the frequency if the on-site potential is hard. In fact, if the dependence of the energy versus frequency presents an extremum, there is a tangent bifurcation at that point. Similarly to the Vakhitov-Kolokolov criterion, our energy criterion is unable to predict oscillatory instability (in our case, caused by Hopf bifurcations). In addition, it is also unable to predict the bifurcations from translational modes, i.e. the one which produce moving breathers, or those bifurcations causing blow-up.

\begin{figure}[tbp]
\begin{center}
\begin{tabular}{cc}
\includegraphics[height=4.5cm]{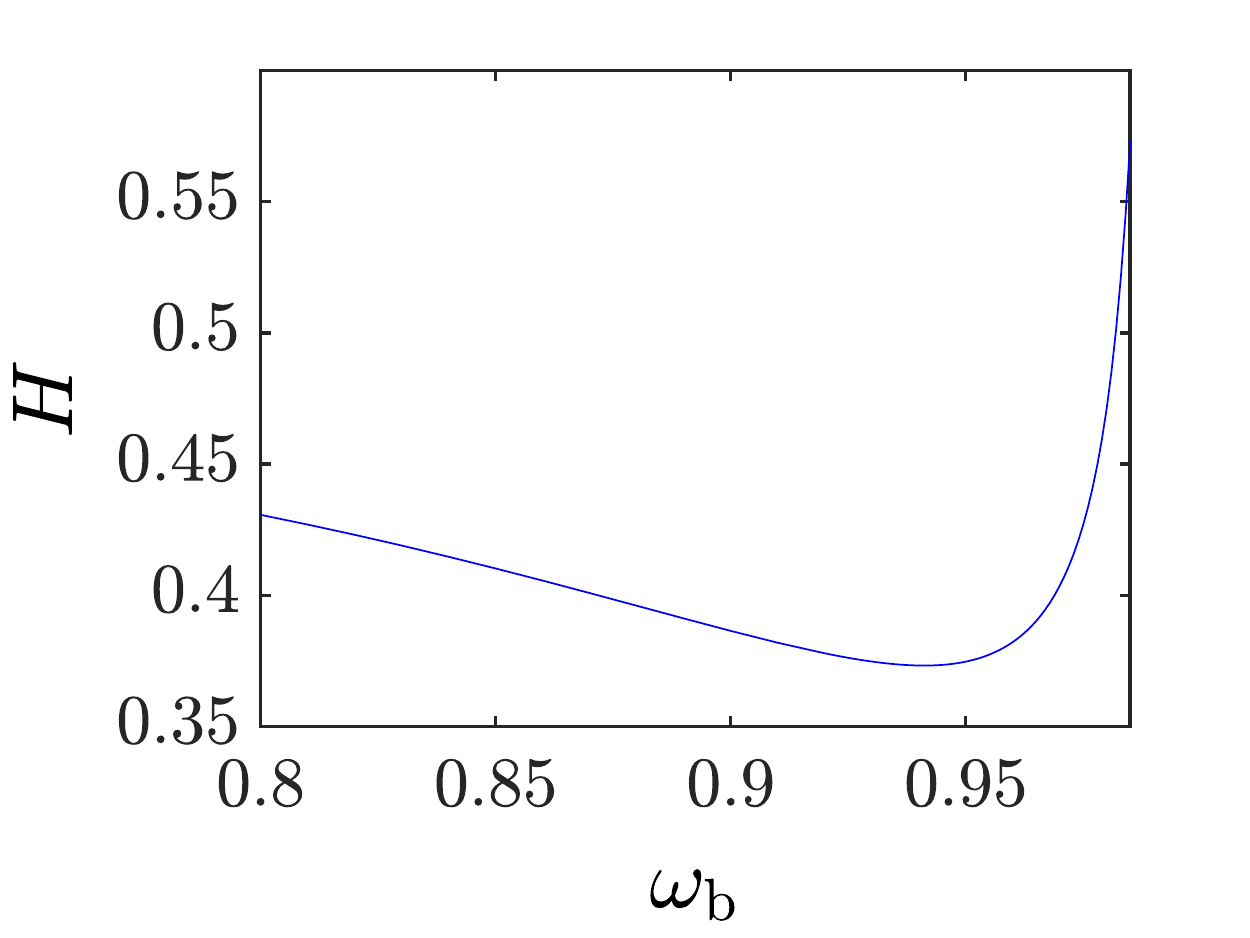} &
\includegraphics[height=4.5cm]{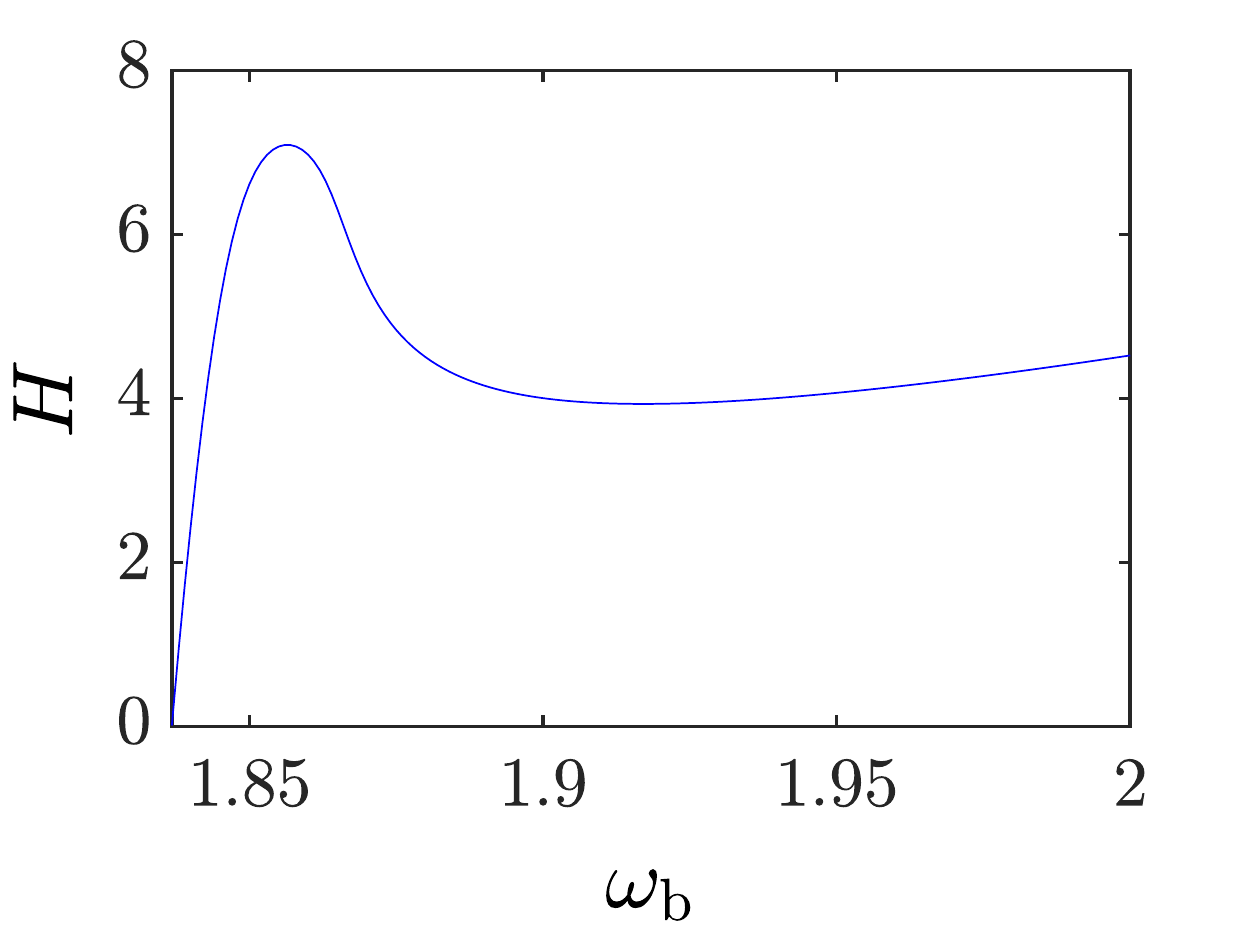} \\[3mm]
\includegraphics[height=4.5cm]{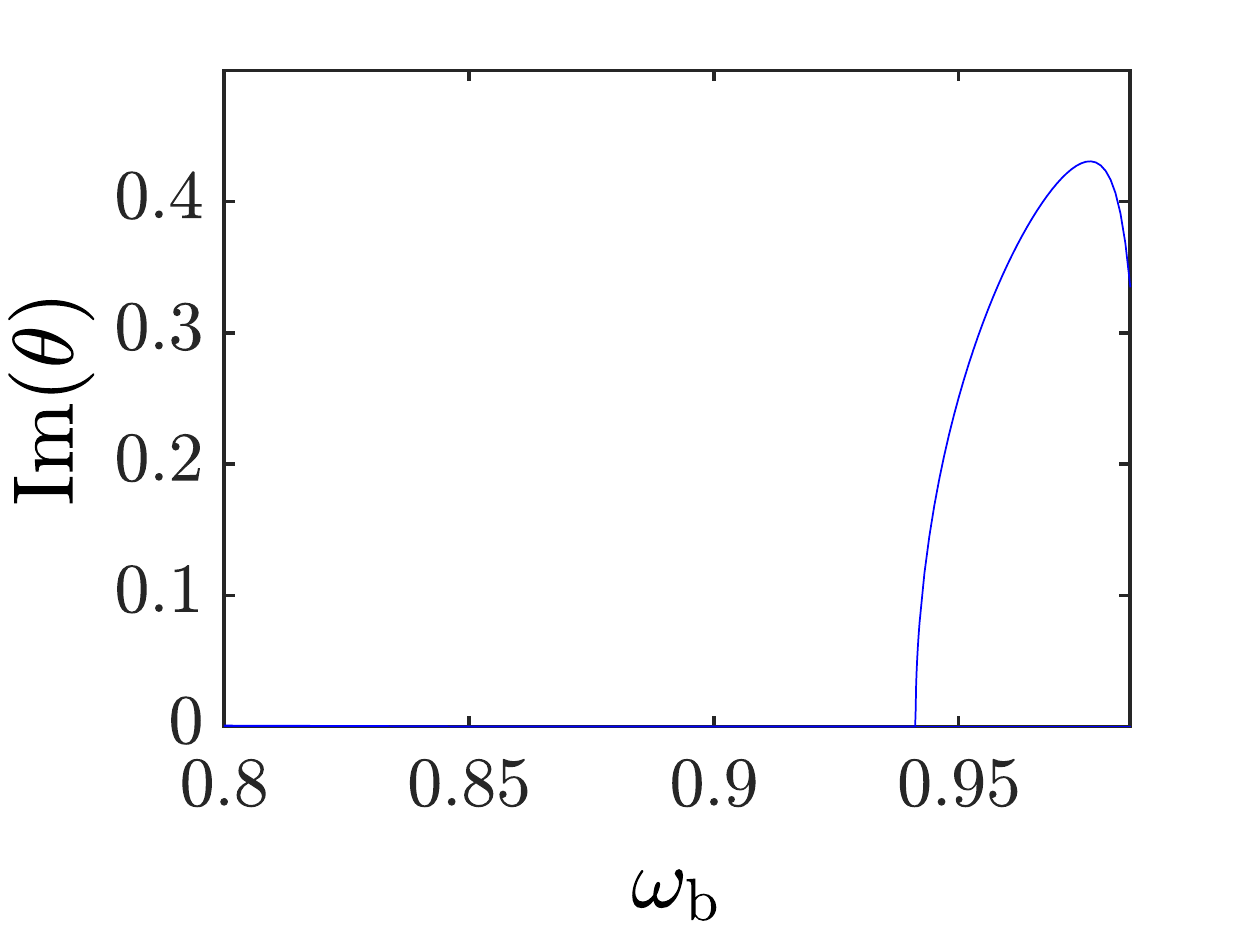} &
\includegraphics[height=4.5cm]{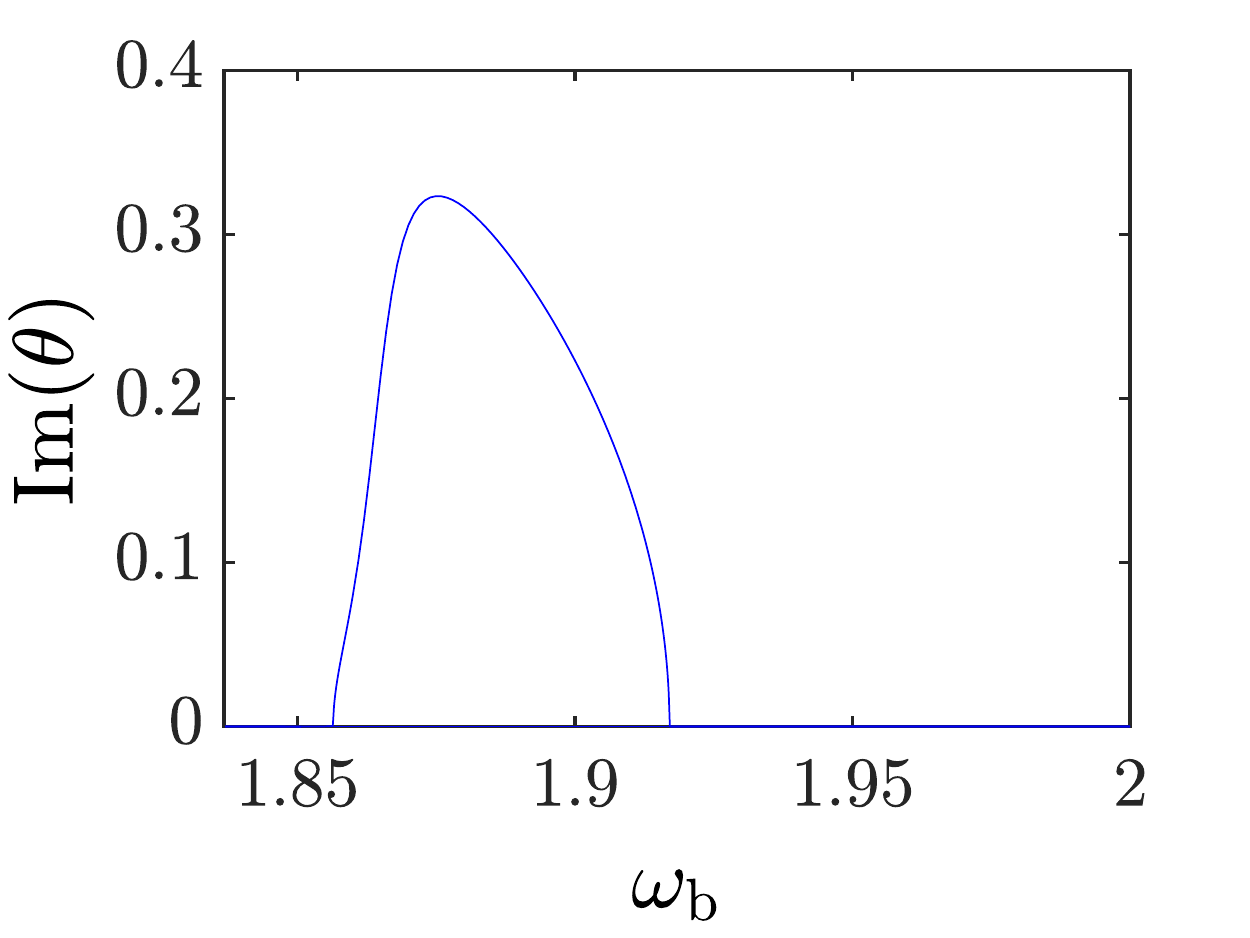}
\end{tabular}
\end{center}
\caption{The top panels show the energy-frequency dependence of discrete breathers in 2D lattices with a soft potential and $C=0.1$ (left panels) and a hard potential and $C=0.3$ (right panels). Bottom panels show the imaginary part of the argument of the Floquet multipliers of these solution families.}
\label{fig:energy}
\end{figure}

In the case of $\phi^4$ potentials, one-dimensional 1-site breathers do not present energy thresholds. But, as mentioned above, the threshold is present for 2D and 3D lattices. This implies an extremum in the energy-frequency dependence and, consequently, a tangent bifurcation that can be predicted by the energy-based criterion of \cite{Energy}. This is shown in Fig.~\ref{fig:energy}.
There, it is clear that for the 2D lattices the change
of monotonicity leads to the corresponding
change of stability.

  The energy criterion has been successfully applied for predicting stability changes in travelling waves (supersonic lattice solitons) in FPUT lattices with soft potentials \cite{Haitao1,Haitao2}. In particular, the connection between
  breathers and traveling waves is an intriguing one: traveling waves
  on a lattice typically return to themselves upon translation by a
  single lattice site. As such, if one considers the operation including
  traveling by one lattice site and then back-shift to the original
  starting point, this operation has a fixed point (if the
  original lattice has a genuine traveling wave). The period of the
  associated periodic orbit is directly associated with the speed
  of the traveling wave (TW) since $\wb=2 \pi s/h$ where $s$ is the speed
  of the wave and $h$ the spacing of the lattice. We thus give a
  very short proof of the theorem for the case of traveling waves
  and, by direct analogy, for DBs.

  Consider the dynamical system:
  $U_t=\left(\begin{array}{c}
  {u} \\
  {p}
\end{array} \right)_t= J \nabla {\cal H}(U)$
where
$J=\left(\begin{array}{cc}
0 & I \\
-I & 0
\end{array} \right)$.
We seek a TW in the form:
$u=u_0(n- s t)=u_0(\xi)$
[solving $-s U_{0,\xi}=J \nabla {\cal H}(U_0)$] and linearize around
it according to: $u(\xi,t)=u_0(\xi)+ \epsilon
W(\xi,t)$ and $p(\xi,t)=p_0(\xi) + \epsilon P(\xi,t)$. Then, the
linearization operator and its adjoint can be found explicitly as:
$\mathcal{M}:= s \partial_{\xi}+J \nabla^2 {\cal H}(U_0), \quad
%  \label{eqn4}
%\end{eqnarray}
%\begin{eqnarray}
\mathcal{M}^*=(- \nabla^2 {\cal H}(U_0) J- s \partial_{\xi})=J\mathcal{M}J$.
Similarly to the breather problem, the time translation invariance
induces a eigenvector and a generalized eigenvector with $0$
eigenvalue of the form: $\partial_{\xi}
U_0$ (the eigenvector) and $\mathcal{M}(-\partial_{s} U_0)=\partial_{\xi}
U_0$ (the equation for the generalized eigenvector).

Then the straightforward proof of the stability theorem is as
follows:
If an extra eigenvector $\tilde{Y}$ with $\lambda=0$ exists, then
it must satisfy
$\mathcal{M} \tilde{Y} = \partial_{s} U_0$ (yielding an additional
generalized eigenvector). This, however, imposes the following
symplectic orthogonality condition:
\[
\begin{split}
0 &= \langle  J \partial_{\xi} U_0, \mathcal{M} \tilde{Y} \rangle =\int (J\partial_{\xi} U_0) \cdot (\partial_{s}U_0) d \xi
= \int \frac{1}{s} \nabla {\cal H} (U_0)  \cdot \frac{\partial U_0}{\partial s} d\xi \\
&= \frac{1}{s}\int \frac{\partial {\cal H}(U_0)}{\partial s}  d \xi
= \frac{1}{s} H'(s).
\end{split}
\]
Consequently, at this critical point the derivative $H'(s)$ must vanish,
and similarly for DBs $H'(\wb)=0$. As explained in~\cite{Energy} for
DBs and~\cite{Haitao1,Haitao2} for TWs, one can take the calculation further
providing estimates of the bifurcating instability-inducing
multipliers on the two sides of the critical point.

\subsection{Nonlinear stability}

As illustrated in~\cite{PanosDmitry} for a wide range of
NLS type models, linear stability is not the last word, as there are cases of e.g. spatially-antisymmetric solitons, which are linearly stable but the dynamics is can be nonlinearly unstable, if it is evolved over sufficiently
long time scales.
In both lattice and continuum NLS modes, it was shown in this work
that spectral stability may be inconclusive if modes of negative
Krein signature exist in the spectrum and if their (nonlinearity
induced) harmonics are in resonance with the continuous spectrum.
In particular, an ordinary differential equation (ODE) was derived suggesting
that for positive energy modes, their energetic content is depleted
due to dispersive wave radiation. However, for negative Krein signature
modes, the reverse process occurs, eventually pumping (over a slow,
power-law in time procedure) the internal mode and ultimately
leading to the demise of the coherent structure due to this
genuinely nonlinear mechanism.

Based in such findings, in \cite{SAM} we explored the nonlinear instability of linearly stable multibreathers. We found that nonlinear instability is possible when an integer multiple of the frequency of an internal (i.e. localized) eigenmode resonates with the phonon band and the Krein signature of such a mode and the phonon arc are  opposite to each other. Notice that the frequency $\Omega$ of the internal eigenmode and its Floquet argument follows the relation $\theta=\pm\Omega T\ \mathrm{mod}\ 2\pi$; consequently, if the coupling constant is small enough, the internal mode has not collided with the phonon arc (in fact, this condition is necessary in order for the breather to be  stable
against potential Hamiltonian Hopf bifurcations),
then $\Omega=\theta/T=\wb\theta/(2\pi)$. In this case, however, a harmonic
of such a mode can be located inside the arcs. If it is the 2nd
harmonic, then the slow growth follows a $t^{-1/2}$ law, if the
3rd, a $t^{-1/4}$ law etc., as derived by the corresponding
ODE~\cite{PanosDmitry,SAM}.

Having in mind the stability properties of multibreathers near the AC limit, the only stable solutions among them are those whose codes $\sigma$ are in phase if the on-site potential is hard and in anti-phase if the potential is soft. As demonstrated in \cite{SAM}, the internal modes that detaches from $\theta=0$ have, in the upper half-circle (i.e. for $\theta\in[0,\pi]$, Krein signature $\kappa=1$ if the potential is soft and $\kappa=-1$ if it is hard. On the other hand, in order to have nonlinear instability, it is needed that the Krein signature of the phonon arcs in the upper half-circle are the opposite to the internal modes. Because of this, as explained at the beginning of Subsection \ref{sec:stabAC}, nonlinear instability can only be possible if $\wo<\wb\leq2\wo$ if the potential is hard and $\frac{2}{2k+1}\wo<\wb<\frac{1}{k}\wo$ with $k\in\mathbb{N}$ if the potential is soft. These conditions were corroborated in the detailed
numerical computations of~\cite{SAM}. In the simpler case of the NLS
and DNLS models, no such conditions need to be imposed, however numerical
computations verified the existence of the instability in~\cite{PanosDmitry}.
Here, we only provide a prototypical case example
of the associated phenomenology in the right panel in Fig.~\ref{fig:stability}.
This shows the particular case of a the $\sigma=\{1,1\}$ multibreather in a hard potential. This solution is linearly stable before the Hopf bifurcation takes place, although presents nonlinear instability as the conditions of the paragraph above are fulfilled. The dynamics eventually manifests this instability
although the latter only arises over extremely long times, much longer
than the linearly unstable cases of the left and middle panels.

\subsection{Dynamics}

In this subsection we will further discuss some
prototypical examples of unstable dynamics stemming from linear and nonlinear instabilities of the (2-site) multibreathers shown previously (especially in Fig.~\ref{fig:stability}).

\begin{figure}
\begin{center}
\begin{tabular}{ccc}
\includegraphics[height=3cm]{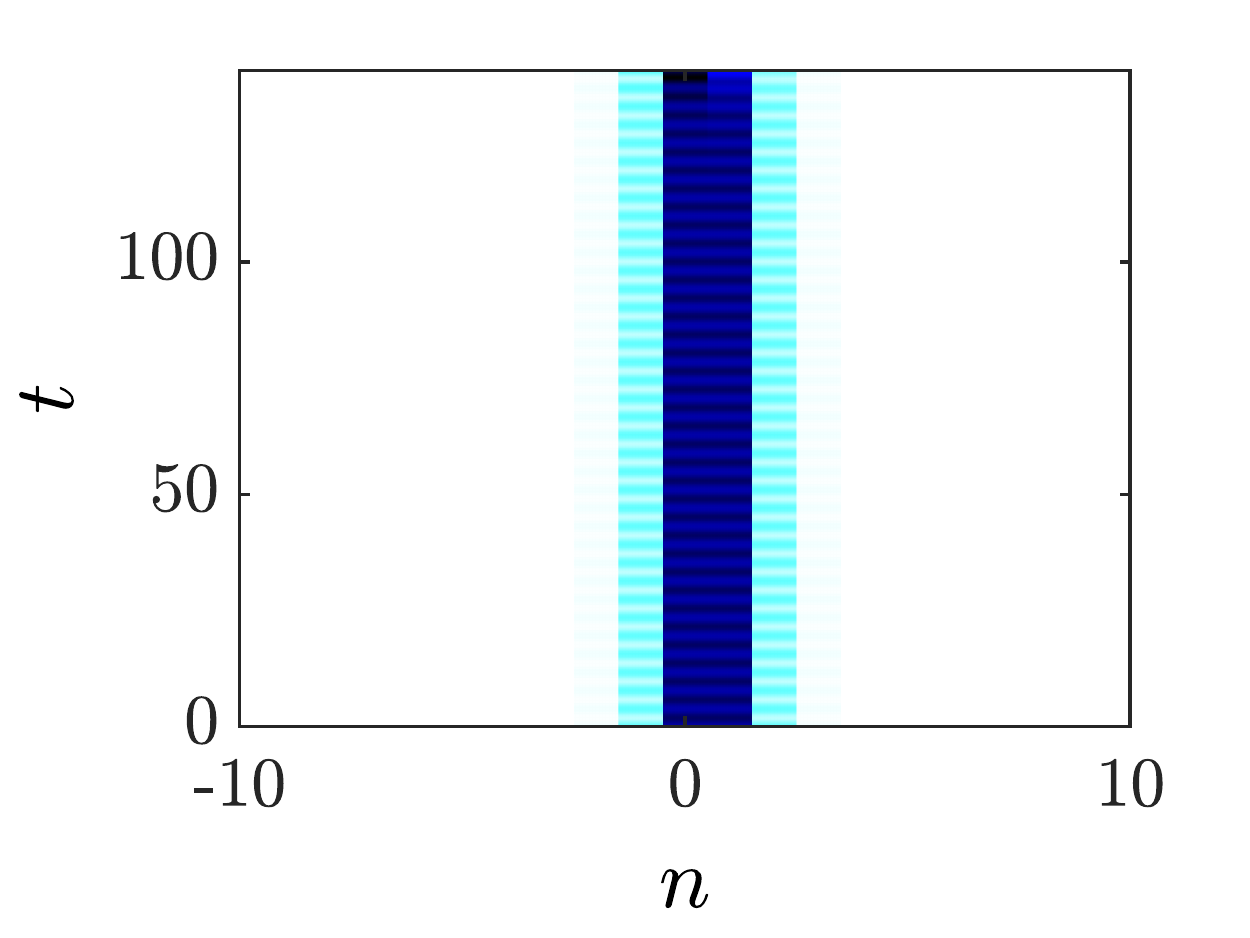} &
\includegraphics[height=3cm]{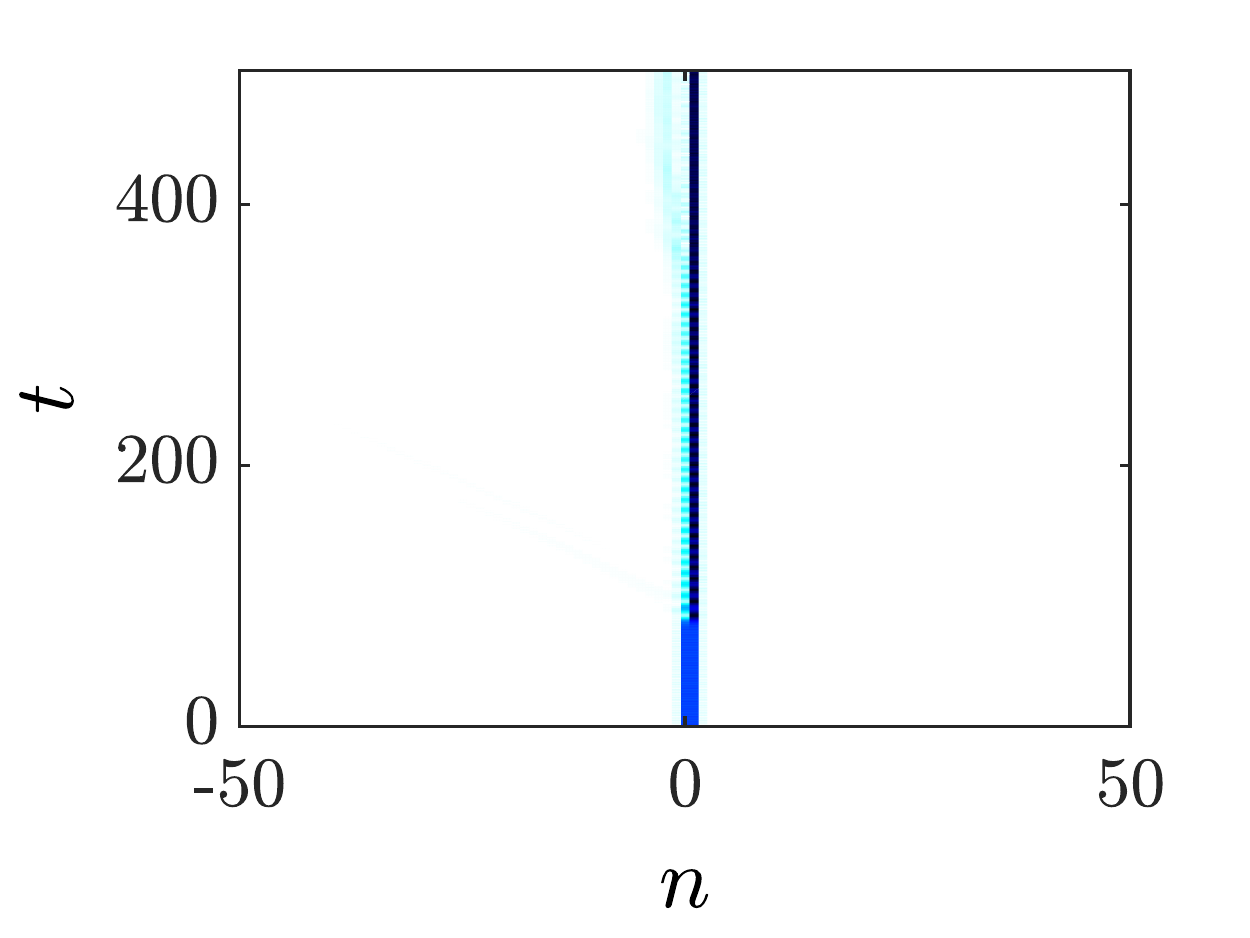} &
\includegraphics[height=3cm]{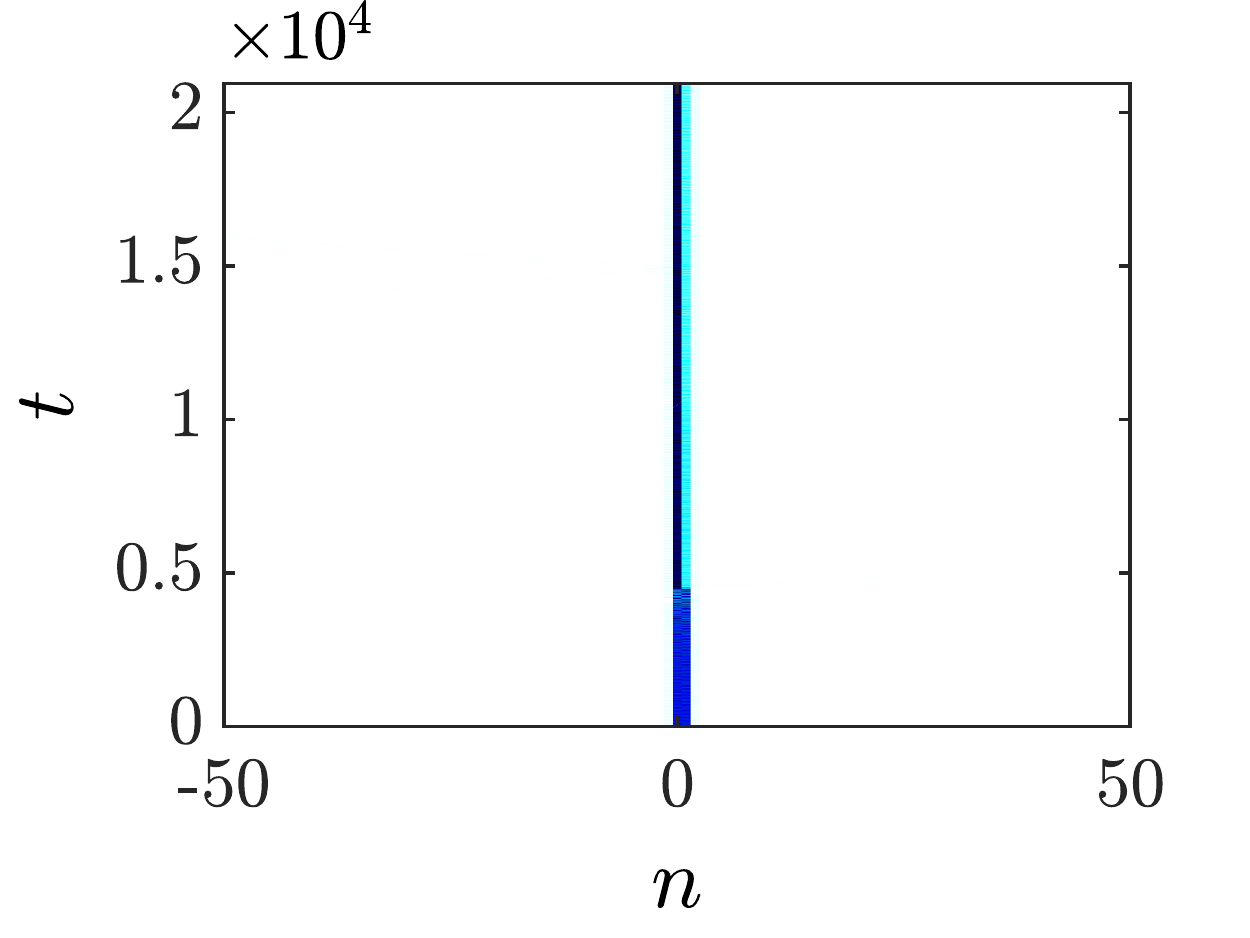} \\[3mm]
\includegraphics[height=3cm]{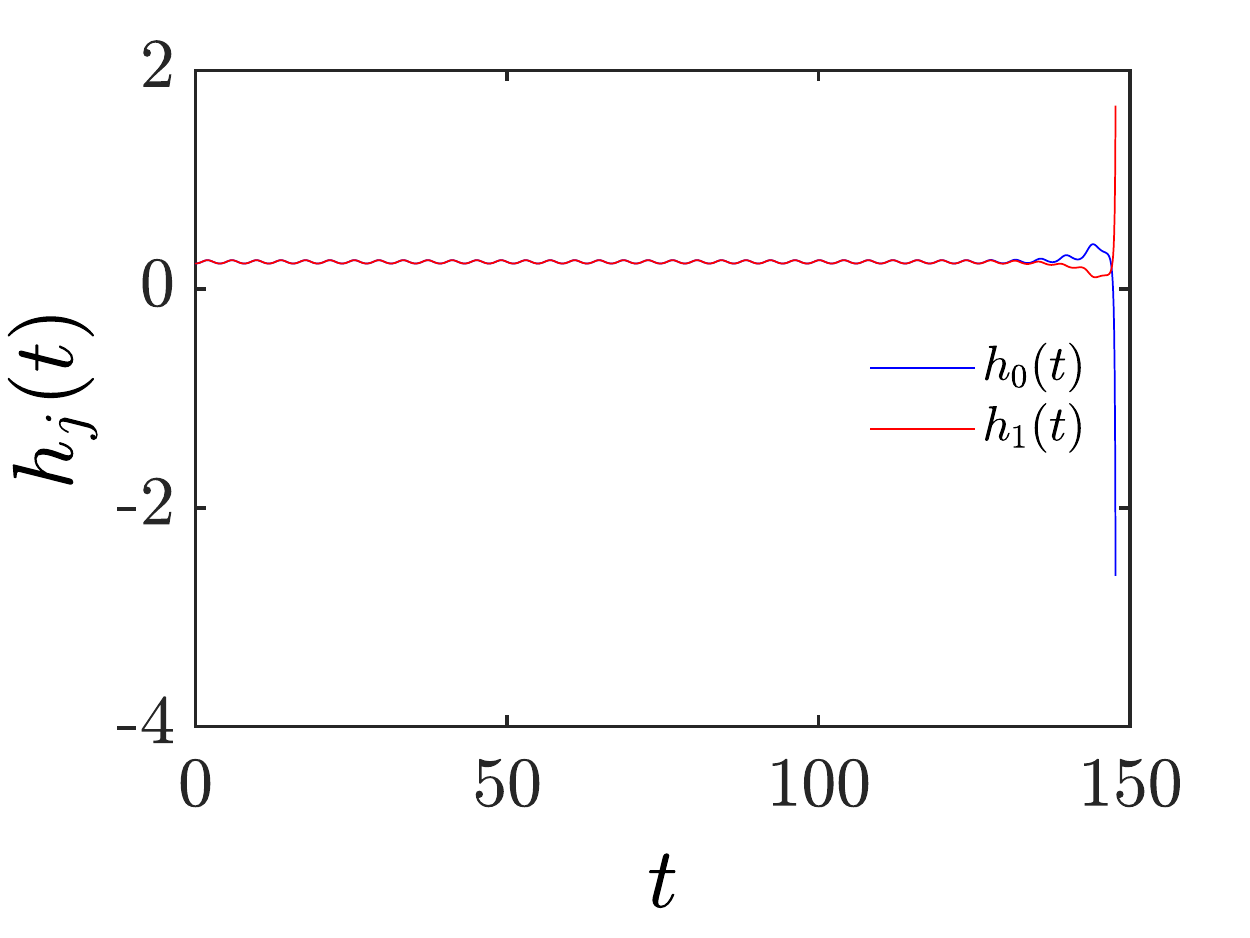} &
\includegraphics[height=3cm]{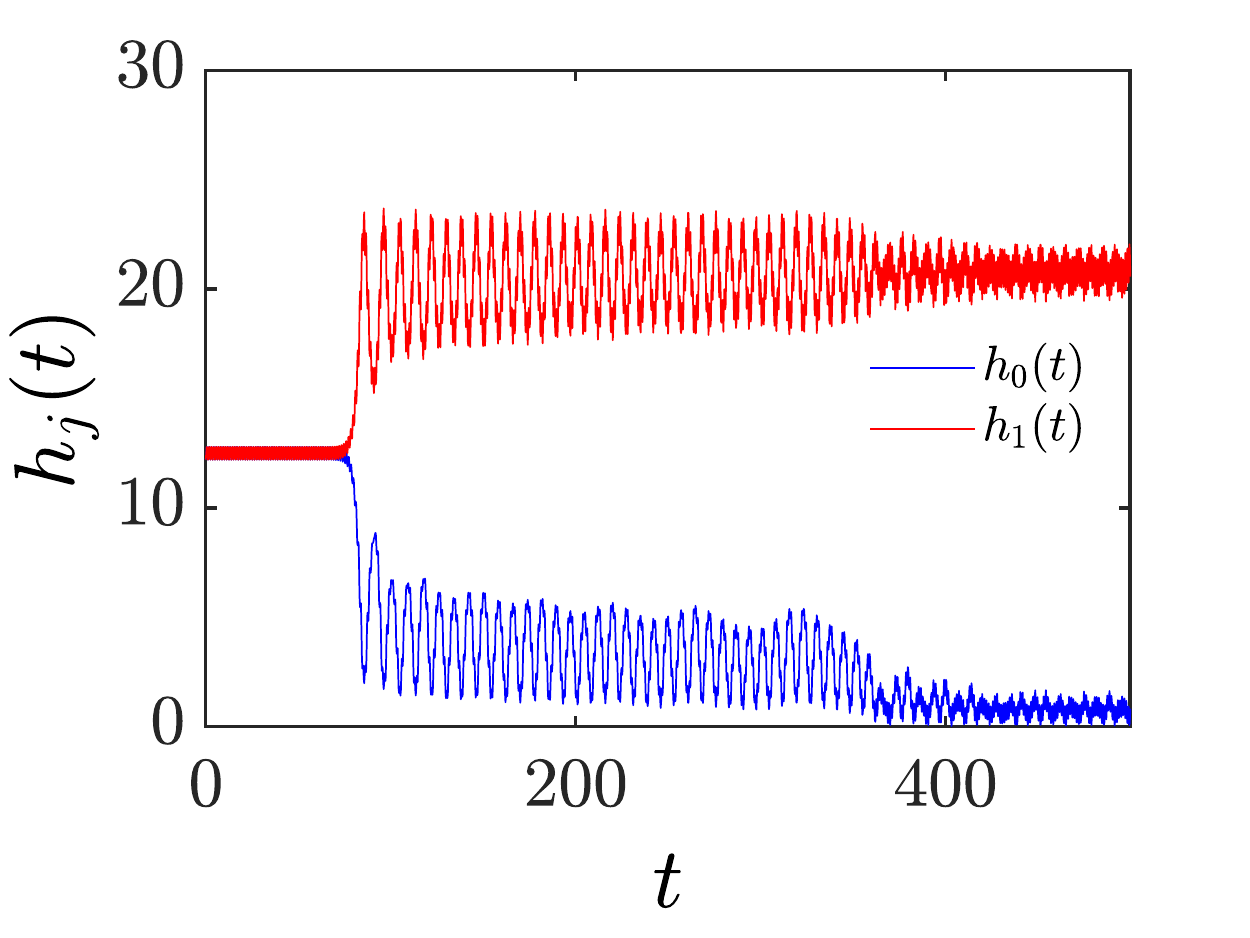} &
\includegraphics[height=3cm]{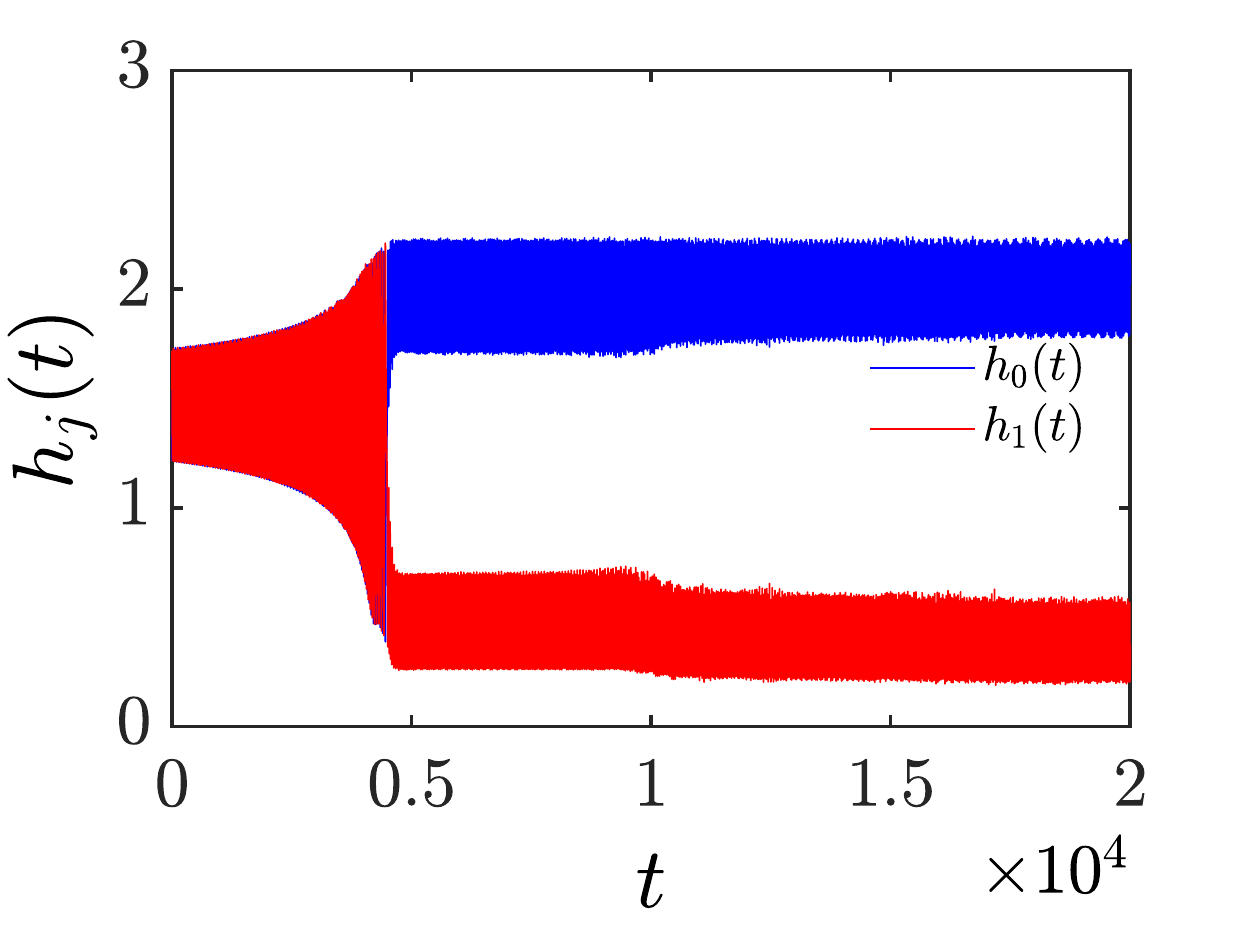} \\[3mm]
\includegraphics[height=3cm]{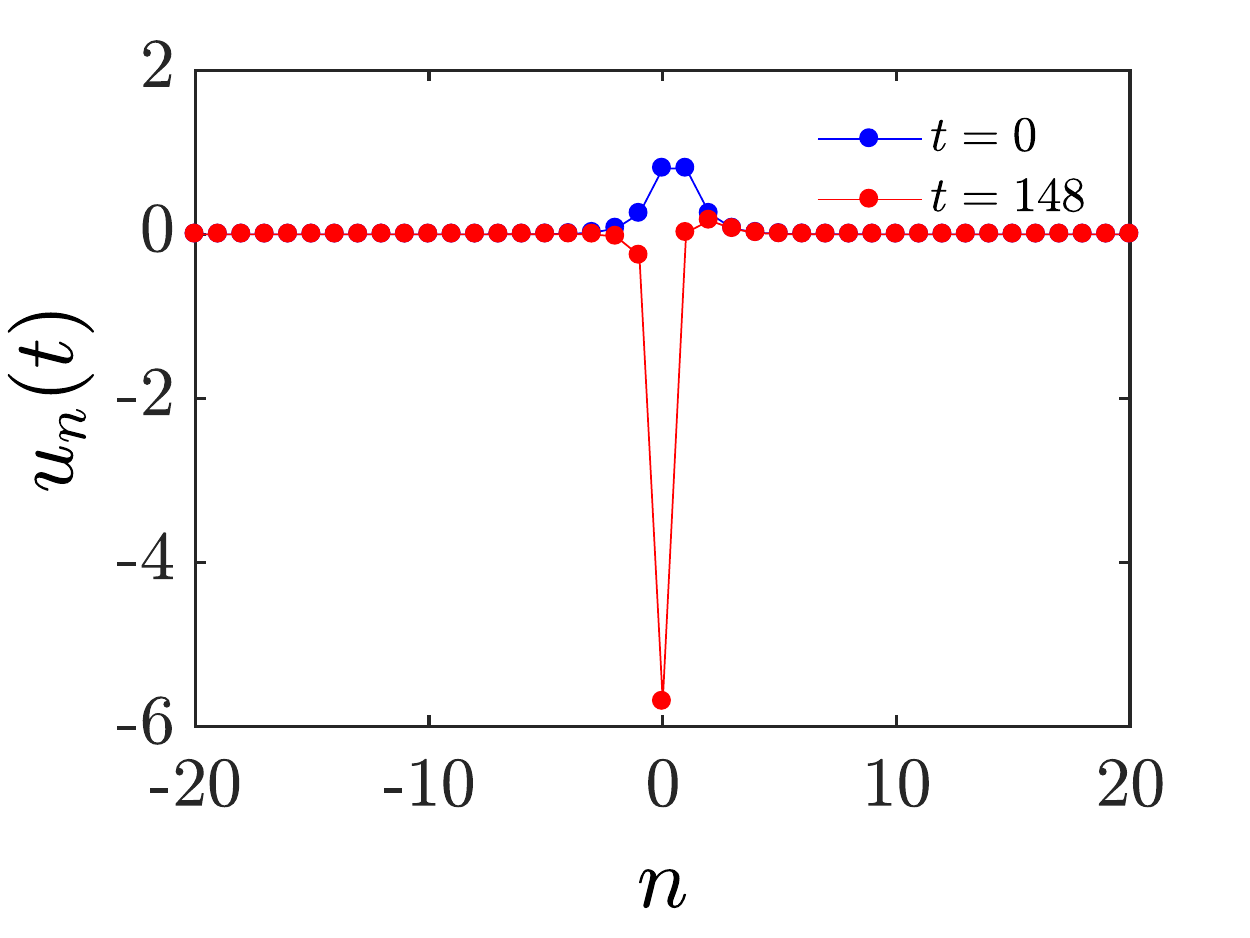} &
\includegraphics[height=3cm]{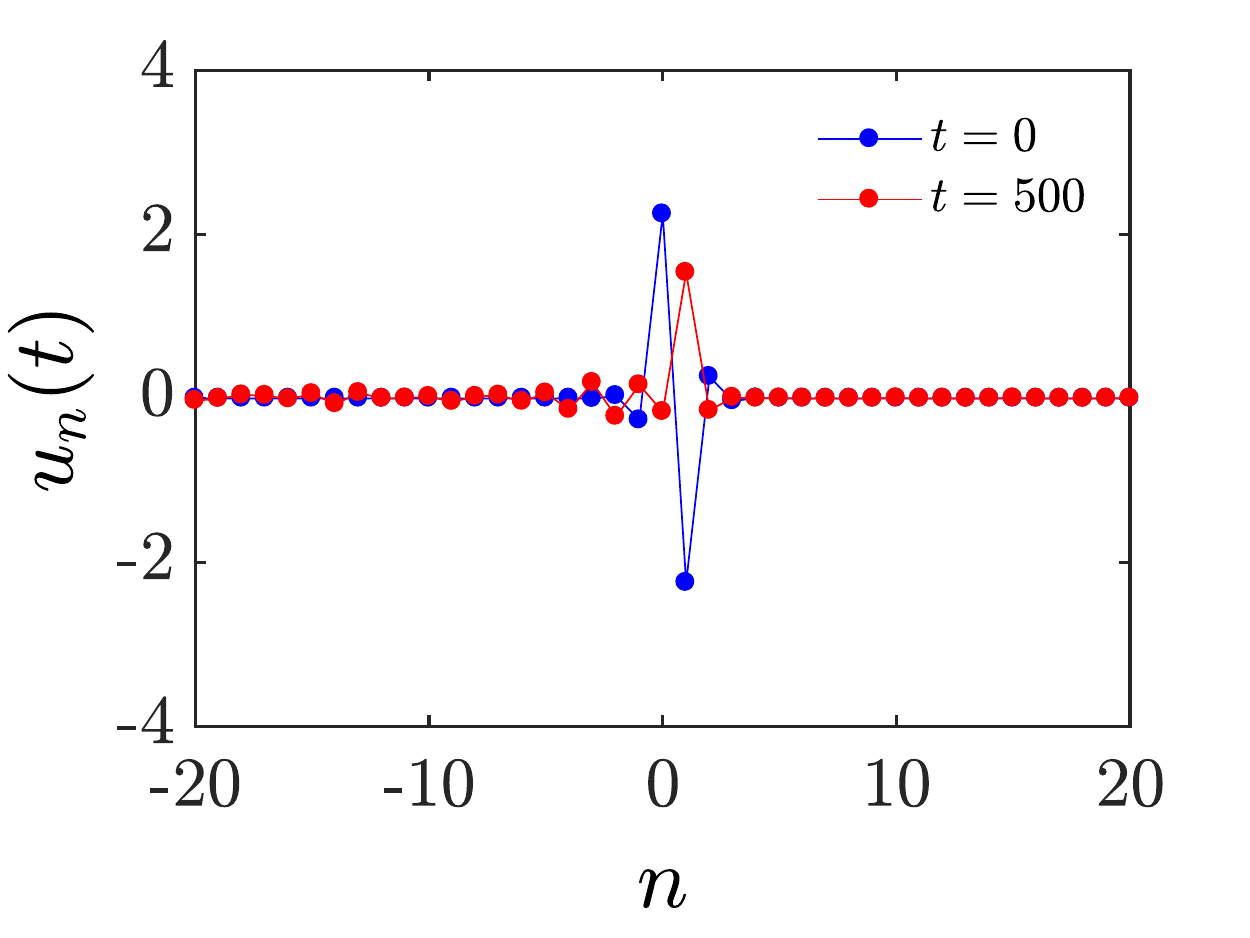} &
\includegraphics[height=3cm]{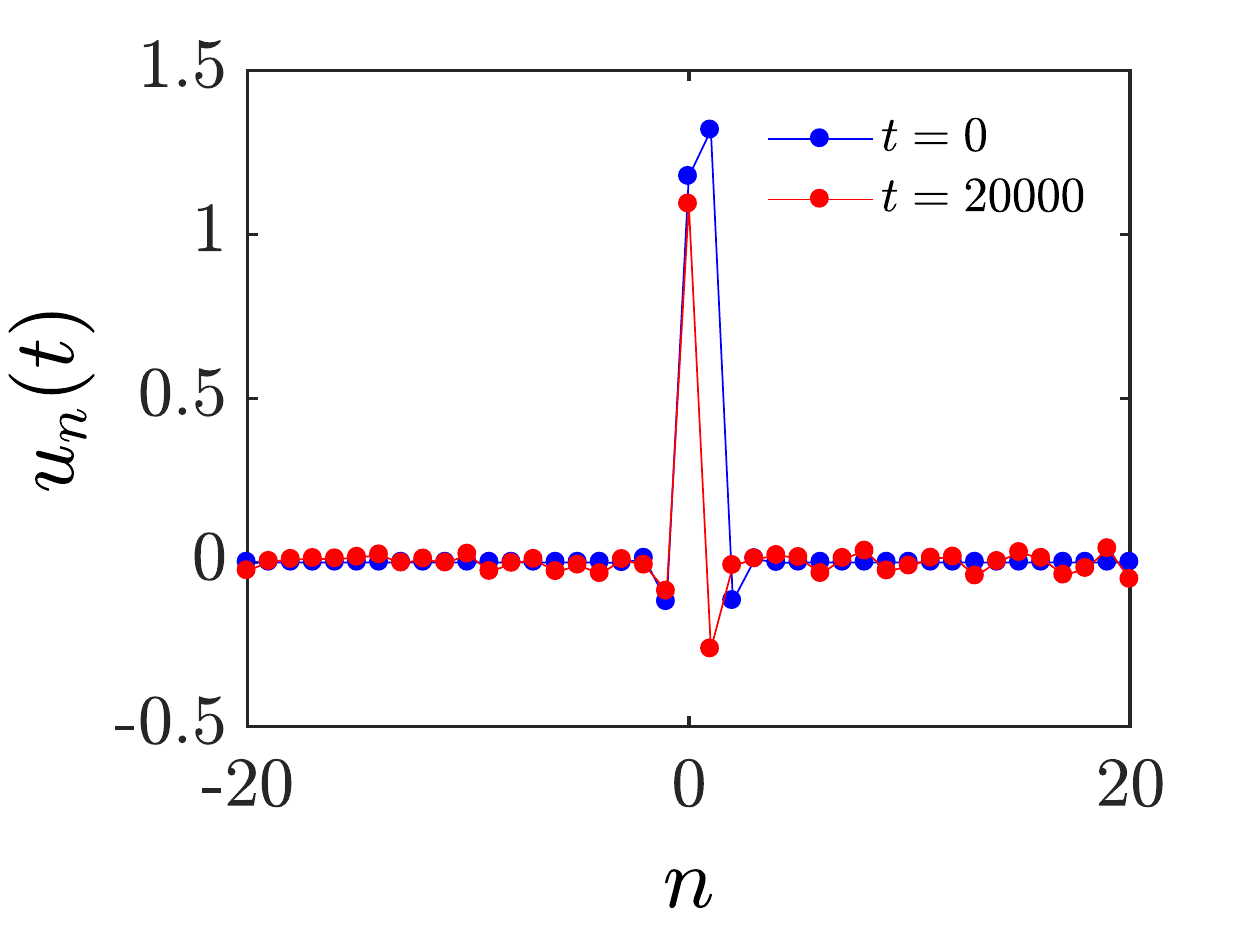}
\end{tabular}
\end{center}
\caption{Dynamical evolution of the unstable configurations displayed in Fig.~\ref{fig:stability}. Top panels show the space-time dependence of the energy density, middle panels depict the energy density evolution of the central sites and bottom panels compare to the profiles of the perturbed breather at the beginning and at the end of the simulation. The multibreathers considered at the figures are: (left panels) $\sigma=\{1,1\}$ in a soft potential with $\wb=0.8$ and $C=0.2$; (central panels) $\sigma=\{1,-1\}$ in a hard potential with $\wb=2.5$ and $C=0.5$; (right panels) $\sigma=\{1,1\}$ in a hard potential with $\wb=1.5$ and $C=0.1$.}
\label{fig:dynamics}
\end{figure}

First of all, we consider linearly unstable multibreathers in the soft $\phi^4$ potential. In that case, the main behavior is a blow-up caused by a phenomenon similar to the escape observed in \cite{Vassos}. Left panels of Fig.~\ref{fig:dynamics} illustrate the blowing-up dynamics observed in an unstable $\sigma=\{1,1\}$ multibreather. Notice that blow-up is caused by both exponential and oscillatory instabilities. I.e., the linearly unstable growth leads the
oscillators to exit the finite height potential barrier and hence
tend to $\pm \infty$.

In the case of hard potentials, the main dynamical features consist of the transformation of the multibreather into a less energetic 1-site breather. In this transformation, some energy is shed from the breather and, in some cases, it can be localized. The central panels of Fig.~\ref{fig:dynamics} shows the dynamics of the $\sigma=\{1,-1\}$ multibreather. It is natural for the configuration to
tend to the site-centered variant as that is generically stable, as we
discussed above.

Finally, the right panels of the figure consider a linearly stable but nonlinearly unstable multibreather with $\sigma=\{1,1\}$ displaying a similar behaviour. The nonlinearly unstable dynamics fulfills the conditions of the theorem of \cite{SAM} as the frequency of the internal mode is $\Omega=0.2073$ and the spectral bands expands in $[0.32,0.5]$. As a result, the second harmonic of the
internal mode lies in the phonon arc, ultimately (at very long times)
slowly feeding the nonlinear growth and eventually leading to the
destruction of the two-site configuration. Notice, however, as also
indicated above, the characteristically longer (by at least an order
of magnitude) time needed for the manifestation of this (nonlinear)
instability.

\section{Some glimpses on other breather features}

Although the core of the present chapter focuses on the stability properties of discrete breathers in $\phi^4$ lattices, we mention in passing
some  developments related to a judicious selection
of aside topics, namely moving breathers and the generation of discrete breathers in lattices with dissipation.

\subsection{Moving breathers}
\label{sec:moving}

As mentioned in the previous section, when an an anti-symmetric eigenmode detaches from the phonon arc, it can reach $\theta=0$ bringing about a tangent bifurcation. Just at $\theta=0$, the mode is marginal and resembles a translational mode, so that a perturbation along it can set the breather into motion. Contrary to continuous breathers, moving discrete breathers generically radiate phonons (see e.g. \cite{MovingCretegny}) and they eventually stop.

Due to the special features of moving breathers, there
does not exist a systematic underlying mathematical theory that can clearly characterize them. Some attempts of defining a Peierls-Nabarro barrier similar to kinks have been performed, but they only seem to work in FPUT lattices close to the continuum limit \cite{PNSepulchre}. In any case, there must be a mechanism alike to Peierls-Nabarro barrier which is related to the existence of tangent bifurcations of the translational mode that detaches from the phonon arc. Because of this, moving breathers can only be observed in lattices for which
the breathers experience such bifurcations. There are only a few reported cases of one-dimensional KG lattices, namely, with Morse, sine-Gordon and double-well potentials \cite{Chen}. We have also been able to generate moving breathers in two-dimensional KG lattices with Morse potential, a result that have not been published yet. Moving breathers (with high mobility) has been observed in two-dimensional lattices with in-plane degrees of freedom modeling e.g. muscovite mica \cite{QiM} which are also known as quodons and, as mentioned in the Introduction, are speculated to play an important role in charge transport properties in such materials.

Moving breathers do not exist in KG lattices with the
$\phi^4$ potential, as the tangent bifurcation of the translational mode does not take place. However, such a bifurcation was observed for KG/FPUT lattices
with on-site and interaction potentials of the hard $\phi^4$ form \cite{Floria}. Such a lattice has been used for modeling micromechanical cantilever arrays \cite{Sato}. In \cite{wedge} we generated moving breathers in such a model and analyze their interaction with geometrical defects.

\subsection{Dissipative lattices}

Most of the experimental findings of discrete breathers have been achieved on lattices with dissipation and external driving, such as micromechanical \cite{Sato}, pendula \cite{Lars} and Josephson junctions \cite{Trias,Ustinov} arrays, nonlinear electrical lattices \cite{Faustino1,Faustino2} or
granular media~\cite{granular2,granular}. Such classes of systems
remain quite popular to this day with numerous variations continuously
arising including, e.g., piecewise-linear systems emulating the
$\beta$-form of the celebrated FPUT lattice~\cite{watanabe},
or electrical systems involving beyond-nearest-neighbor
interactions~\cite{larsnew}.

As demonstrated in \cite{Sepulchre2}, discrete breathers in dissipative lattices can also exist away from the anticontinuum limit. Contrary to Hamiltonian lattices, there are no resonances with phonons (the spectrum
of plane wave excitations is pushed to the left half
of the complex spectral plane); as a result such states are
now potential attractors of the system. The work of \cite{Marin2} shows the complex phenomenology that is observed in driven and damped Frenkel--Kontorova lattices. If the system is driven with a frequency $\wb$, discrete breather solutions acquire the same frequency as the driving force. In general, all the lattice sites oscillate with the same frequency; however, we have found an electrical lattice where subharmonic resonance emerges (that is, the excited sites of the breather oscillate with half of the frequency of the low amplitude sites) \cite{Faustino3}. Recently, multistable variations of pendula, potentially applicable
to SQUID metamaterials, have been
manifested as potential sources of more complex breathing patterns
such as the celebrated chimera states~\cite{hitza}.

Discrete breathers have been studied in driven and damped one-dimensional KG lattices with hard $\phi^4$ potentials in \cite{David}. Such a lattice is defined by equation:

\begin{equation}
    \ddot{u}_n+\alpha\dot u_n+V'(u_n)+C(2 u_n-u_{n+1}-u_{n-1})=F_n(t)+\eta_n(t)
\end{equation}
with $F_n(t)$ being a periodic function of frequency $\wb$ and $\eta_n(t)$ is a Gaussian white noise with zero mean and autocorrelation $<\eta_n(t)\eta_m(t')>=2D\delta_{nm}\delta(t-t')$. In the deterministic case ($D=0$) and for
a staggered driving of the form $F_n(t)=(-1)^nf\sin(\wb t)$, the phenomenology is quite simple: at given damping and frequency, discrete breathers only exist above a threshold $f_\mathrm{th}$. However, if the noise is introduced in the lattice, there are two interesting phenomena: if the driving amplitude is suprathreshold ($f>f_\mathrm{th}$), breathers with frequency $\wb$ can be generated even if the initial condition is uniform; if the driving amplitude is subthreshold ($f<f_\mathrm{th}$), breathers are still produced by the concerted action of noise and the driving force, in a way that noise, on the one hand, enables system transitions between the uniform and the coherent localized states and, on the other hand, destroys any degree of order of the system if its amplitude is large enough: in other words, we are dealing with a stochastic resonance phenomenon.

\section{Outlook and future directions}

From the above discussion, it is clear that the theme of
discrete breathers is one that is gradually maturing and
emerging in a wide range of applications and a diverse
array of systems. In particular, we are gradually
departing from the simplest nearest-neighbor scenarios
of either just KG or just FPUT types and moving on to
a new, more elaborate phase where systems
can be designed with beyond-nearest~\cite{larsnew} and even long-range
interactions~\cite{cho} and also with multiple and potentially
competing~\cite{kimura} interactions, or with ones
that are progressively more amenable to analytical
considerations~\cite{watanabe}. This suggests that
there is a significant need for further theoretical
and computational developments to support the corresponding
emerging experimental platforms.

While the early stages of development of DBs favored analytical
proofs of existence and associated techniques of numerical
existence and stability, subsequent ones favored a more
systematic exploration of spectral properties and an attempt
to classify the different multibreathers and offer systematic
guidelines about when they may be expected to be dynamically robust.
In this Chapter, we summarized some of this systematic effort
in the previous decade and some of its crystallized results
and convergence of different methods over the past few years.
More recently, further tools have arisen in probing spectral
and dynamical features of DBs. Among others, we have explored
and summarized here the energy-versus-frequency monotonicity
criteria and how they relate to linear instabilities and given connections
between these and the stability of traveling waves in lattices.
Additionally, we have warned the reader against the naive expectation
that spectral stability is the full story, presenting case
examples where this fails to be true due to the nonlinear
instability of internal modes with opposite Krein signature
than that of the phonon arcs. The slow, power-law nature
of the latter instabilities, as opposed to the exponential
growth of linear instabilities was highlighted. Lastly,
some possibilities of further developments towards moving
breathers or dissipative lattices were briefly touched upon.

Clearly, there is need for further theoretical systematics.
Many of the relevant points were brought up in parts of
our discussion. Understanding the stability of phase-shift
multibreathers and vortex breathers in higher-dimensional
systems is an important open topic. Carrying out the
associated stability computations to higher order is
particularly relevant. Recent work, in fact, brings up
the possibility that relevant solutions may fail to exist
at higher orders in some important case examples~\cite{tiziano}.
Studying also long-range interactions may bring about
surprises and produce gaps in the spectrum where novel DBs may exist,
as per the recent work of~\cite{doi}. Again, this is a topic meriting
further exploration. The study of systems with nontrivial tails
(nanoptera) and the examination of whether the stability features/criteria
presented herein apply to them is also an open topic.
The same holds true for systems with external drive and damping:
can we offer some guidelines to characterize their stability
characteristics, suitably adapting what we know in the more structured
Hamiltonian cases or perhaps not~?
Plus then there are topics which, while touched
upon, still seem fairly poorly understood or wide open for new insights:
among them moving breathers, or quasi-periodic solutions and their
existence and stability, as well as the role of DBs in
asymptotic dynamics and thermalization (see, e.g.,~\cite{flachtherm}
for a recent summary in the discrete NLS case) are only some that come to mind.
In summary, discrete breathers may have matured but have many
more challenges to offer for the years to come both at the fundamental,
at the computational and at the experimental level...

\begin{acknowledgement}
This material is based upon work supported by the National Science Foundation under Grant No. DMS-1809074 (P.G.K.). J.C.-M. thanks financial support from MAT2016-79866-R project (AEI/FEDER, UE). P.G.K. also gratefully acknowledges support from the US-AFOSR under Grant No. FA9550-17-1-0114.
\end{acknowledgement}

\end{document}